\pdfoutput=1
\PassOptionsToPackage{table}{xcolor}

\documentclass[journal]{IEEEtran}

\usepackage[T1]{fontenc}
\usepackage[utf8]{inputenc}
\usepackage{textcomp}
\usepackage{cite}
\usepackage[pdftex]{graphicx}
\usepackage{amsmath}
\usepackage[cmintegrals]{newtxmath}
\usepackage{trfsigns}
\usepackage{upgreek}
\usepackage{diagbox}
\usepackage[normalem]{ulem}
\usepackage{tikz}
\usetikzlibrary{calc,shapes,arrows,positioning,arrows.meta,bending,automata,quotes,angles,matrix,decorations.pathreplacing,backgrounds}

\usepackage{tikz-3dplot}
\usetikzlibrary{3d}

\tikzstyle{sum} = [thick,draw, circle,inner sep=-1pt,outer sep=0pt,font=\normalsize]

\pgfmathsetmacro{\blockwidth}{30}
\pgfmathsetmacro{\blockheight}{17}

\tikzstyle{comblock} = [thick,black,fill=white,draw=black, rectangle,minimum height=\blockheight pt,minimum width=\blockwidth pt]
\tikzstyle{comtriag} = [regular polygon, regular polygon sides=3,
              draw=black, black,  text width=0.90em,
              inner sep=0.9mm, outer sep=0mm,
              shape border rotate=-90]

\tikzstyle{input} =  [coordinate]
\tikzstyle{output} = [coordinate]

\tikzset{%
Double/.style={%
    to path={%
      ($(\tikztostart)!2pt!90:($(\tikztotarget)!5pt!(\tikztostart)$)$) -- ($($(\tikztotarget)!2.3pt!(\tikztostart)$)!2pt!270:(\tikztostart)$)
      ($(\tikztostart)!2pt!270:($(\tikztotarget)!4pt!(\tikztostart)$)$) -- ($($(\tikztotarget)!2.3pt!(\tikztostart)$)!2pt!90:(\tikztostart)$)
      ($($(\tikztotarget)!5pt!(\tikztostart)$)!5pt!90:(\tikztostart)$)
       .. controls
          ($($(\tikztotarget)!3pt!(\tikztostart)$)!3pt!90:(\tikztostart)$) and
          ($(\tikztotarget)!0.1pt!(\tikztostart)$)
       .. (\tikztotarget)
       .. controls
          ($(\tikztotarget)!0.1pt!(\tikztostart)$) and
          ($($(\tikztotarget)!3pt!(\tikztostart)$)!3pt!270:(\tikztostart)$)
       ..
     ($($(\tikztotarget)!5pt!(\tikztostart)$)!5pt!270:(\tikztostart)$)
    }
  }
}

\usepackage[exponent-product = \times]{siunitx}
\usepackage{graphicx}

\usepackage{cleveref}
\usepackage{mathtools}
\usetikzlibrary{patterns}

\usepackage{pgfplots}
\pgfplotsset{compat=1.14}
\usepackage{pgfplotstable}

\usepackage{booktabs}
\usepackage{trfsigns}
\usepackage[b]{esvect}
\usepackage{blkarray}
\usepackage{multirow,array}

\usepackage[linesnumbered,ruled]{algorithm2e}

\SetCommentSty{mycommfont}

\usepgfplotslibrary{fillbetween}

\makeatletter
\newcommand*\bigcdot{\mathpalette\bigcdot@{.5}}
\newcommand*\bigcdot@[2]{\mathbin{\vcenter{\hbox{\scalebox{#2}{$\m@th#1\bullet$}}}}}
\makeatother

\definecolor{mycolor1}{HTML}{3070b3}%
\definecolor{mycolor2}{HTML}{A2AD00}
\definecolor{mycolor3}{HTML}{E37222}
\definecolor{mycolor4}{HTML}{DAD7CB}

\DeclareMathOperator*{\T}{T}

\DeclareMathOperator*{\E}{E}
\DeclareMathOperator{\sinc}{sinc}

\newcommand{\moment}[1]{\ensuremath{\mathrm{\mu}_{4}}}

\newcommand{\dimC}[1]{\ensuremath{\in \mathbb{C}^{#1}}}
\newcommand{\dimR}[1]{\ensuremath{\in \mathbb{R}^{#1}}}
\newcommand{\covm}[2]{\ensuremath{\boldsymbol{\mathbf{#1}}_{{#2}}}}

\newcommand{\mident}[1]{\ensuremath{\mathrm{\textbf{I}}_{#1}}}
\newcommand{\mnull}[1]{\ensuremath{\mathrm{\textbf{0}}_{#1}}}

\newcommand{\GaussDist}[3]{\ensuremath{\mathcal{N}\!\left(#1;\,#2\,,#3\right) }}

\newcommand{\sicindex}{t} 
\newcommand{\siclength}{N}

\newcommand{\gibbsvec}{\widetilde{\mathbf{b}}}
\newcommand{\gibbsscalar}{\widetilde{b}}
\newcommand{\Gibbsvec}{\widetilde{\mathbf{B}}}
\newcommand{\Gibbsscalar}{\widetilde{B}}

\newcommand{\bm}[1]{\ensuremath{\boldsymbol{\mathbf{#1}}}}

\definecolor{mycolor2}{RGB}{196,7,27}
\definecolor{mycolor1}{RGB}{0,101,189}
\definecolor{bblack}{RGB}{088,088,090}
\definecolor{oorange}{RGB}{255,180,000}
\definecolor{mycolor3}{RGB}{255,180,000}
\definecolor{mycolor4}{HTML}{3A7E55}
\definecolor{mycolor5}{HTML}{3A7E55}

\tikzstyle{recty} = [draw, rectangle, scale=1.2,font=\normalsize,minimum height=1.5em,minimum width=3em]

\usepackage{graphicx}
\usepackage{subcaption}
\usepackage{etoolbox}
\usepackage{comment}

\definecolor{mycolor1}{HTML}{0072bd}%
\definecolor{mycolor1dark}{HTML}{0b5b8f}%

\definecolor{mycolor2}{rgb}{1.25000,0.32500,0.09800}%
\definecolor{mycolor3}{rgb}{0.92900,0.69400,0.12500}%
\definecolor{mycolor4}{HTML}{453645}

\definecolor{mycolor5}{HTML}{27ae60}%

\definecolor{mycolor5bright}{HTML}{0A9396}%
\definecolor{mycolor1bright}{HTML}{E9D8A6}%
\definecolor{mycolor7bright}{HTML}{EE9B00}%
\definecolor{mycolor6bright}{HTML}{AFCBFF}%
\definecolor{mycolor5dark}{HTML}{166337}%
\definecolor{mycolor6}{HTML}{FF0000}%

\definecolor{mycolor6dark}{HTML}{961717}%
\definecolor{mycolorlightorange}{HTML}{FFD28E}%
\definecolor{mycolorcube}{HTML}{7C7481}%

\newcommand{\gridfigurewidth}{7.4cm}
\newcommand{\gridfigureheight}{4cm}

\definecolor{mycolor7}{HTML}{0ae4fc}

\pgfplotsset{
   compat=1.15,
   PAM/.style={
    mycolor1,
    line width=0.5pt,
   },
   ASK/.style={
    mycolor6,
    line width=0.5pt,
   },
   ASKGibbs/.style={
    mycolor4,
    densely dotted,
    line width=0.90pt,
   },
    QAM/.style={
       mycolor5,
       line width=0.5pt,
    },
    SP/.style={
        densely dashed,
    },
    TDRC/.style={
       solid,mark=*,mark size=2pt,mark options={draw=white,solid,line width=0.5pt}
    },
    TDRC_ICI/.style={
       solid,mark=diamond*,mark size=2.7pt,mark options={draw=white,solid,line width=0.5pt}
    },
    UB/.style={
       line width=0.85pt,
       mark=none, 
       densely dotted
    },
     LB/.style={
       line width=0.85pt,
       mark=none, 
       densely dash dot
    },
    PAM_L30km/.style={
       mark=triangle*, 
       mark size=4pt,
       mark options={draw=white,solid,line width=0.4pt}
    },
    ASK_L30km/.style={
       mark=square*, 
       mark size=2.5pt,
       mark options={draw=white,solid,line width=0.4pt}
    },
    QAM_L30km/.style={
       mark=*, 
       mark size=2.4pt,
       mark options={draw=white,solid,line width=0.4pt}
    },
    FERGeneralStyle/.style={
      width=\gridfigurewidth,
      height=\gridfigureheight,
      scale only axis,
      xmin=-5,
      xmax=13,
      xlabel style={font=\color{white!15!black}},
      xlabel={SNR $\text{[dB]}$},
      yminorticks=true,
      ylabel style={font=\color{white!15!black}},
      ylabel={FER},
      axis background/.style={fill=white},
      xmajorgrids,
      ymajorgrids,
      yminorgrids,
      xminorgrids,
      xtick distance={0.5},
      legend style={legend cell align=left, font=\scriptsize, align=left, draw=white!15!black,legend pos=north east
      },
    },
    MIGeneralStyle/.style={
      width=\gridfigurewidth,
      height=\gridfigureheight,
      at={(0.968in,0.544in)},
      scale only axis,
      minor y tick num=9,
      yminorticks=true,
      xmajorgrids,
      ymajorgrids,
      yminorgrids,
      xmin=-5,
      xmax=13,
      xlabel style={font=\color{white!15!black}},
      xlabel={SNR [dB]},
      ymin=0,
      ylabel style={font=\color{white!15!black}},
      ylabel={bpcu},
      axis background/.style={fill=white},
      xmajorgrids,
      ymajorgrids,
      xminorgrids,
      xtick distance={2},
      legend style={legend cell align=left, font=\scriptsize, align=left, draw=white!15!black,legend pos=north west},
      }
}

\tikzset{
    arr/.style={latex-latex,very thick},
    arrsingle/.style={latex-,thick},
    shadowed/.style={
    preaction={transform canvas={shift={(-0.3pt,+0.3pt)}},draw=white,thick}},
    shadowedb/.style={
    preaction={transform canvas={shift={(0.3pt,-0.3pt)}},draw=black,very thick}},
    MyEl/.style={
       line width = 0.8pt,
       densely dash dot,
       black!80!gray
    },
    opacitylabel/.style={
       fill=white, fill opacity=0.7,text opacity=1, draw opacity=1
    },
}

\begin{document}
\title{Successive Interference Cancellation for Bandlimited Channels with Direct Detection}

\author{Tobias Prinz,~\IEEEmembership{Member,~IEEE,} Daniel Plabst, 
Thomas Wiegart,~\IEEEmembership{Graduate Student~Member,~IEEE,}\\
Stefano Calabr\`o, Nobert Hanik,~\IEEEmembership{Senior~Member,~IEEE,} and Gerhard Kramer,~\IEEEmembership{Fellow,~IEEE} 
\thanks{Date of current version \today. Accepted to \emph{IEEE Transactions on Communications} on November 17, 2023. (\emph{Tobias Prinz and Daniel Plabst are co-first authors.)}}
\thanks{Tobias Prinz, Daniel Plabst, Thomas Wiegart, Norbert Hanik, and Gerhard Kramer are with the Institute for Communications Engineering,
School of Computation, Information, and Technology, Technical University of Munich, 80333 Munich, Germany (e-mail: tobias.prinz@tum.de, daniel.plabst@tum.de, thomas.wiegart@tum.de, norbert.hanik@tum.de, gerhard.kramer@tum.de).}
\thanks{Stefano Calabr\`o is with the Huawei Munich Research Center, 80992 Munich, Germany (e-mail:
stefano.calabro@huawei.com).}
}

\maketitle

\begin{abstract}
The maximum information rates for bandlimited channels with direct detection are achieved with joint detection and decoding (JDD), but JDD is often too complex to implement. Two receiver structures are studied to reduce complexity: separate detection and decoding (SDD) and successive interference cancellation (SIC). For bipolar modulation, frequency-domain raised-cosine pulse shaping, and fiber-optic channels with chromatic dispersion, SIC achieves rates close to those of JDD, thereby attaining significant energy gains over SDD and intensity modulation. Gibbs sampling further reduces the detector complexity and achieves rates close to those of the forward-backward algorithm at low to intermediate signal-to-noise ratio (SNR) but stalls at high SNR. Simulations with polar codes, higher-order modulation, and multi-level coding confirm the predicted gains.
\end{abstract}

\begin{IEEEkeywords}
Capacity, direct detection, Gibbs sampling, information rate, successive interference cancellation.
\end{IEEEkeywords}

\maketitle

\section{Introduction}
\IEEEPARstart{S}{hort-reach} fiber-optic communication systems often use direct detection (DD) devices with one photodiode (PD) per wavelength that outputs the intensity of the received optical signal~\cite{chagnon_optical_comms_short_reach_2019}. DD allows easy reconstruction of the transmitted data when using intensity modulation (IM) and symbol-rate sampling~\cite{dissanayake_comparison_ofdm_imdd_2013,chen_performance_analysis_ofdm_imdd_2012,mecozzi_imdd_capacity_optical_amp2001}.
The paper~\cite{mecozzi_capacity_amdd_2018} showed that the DD capacity increases by oversampling, i.e., sampling faster than the symbol rate. In fact, if the dominant noise is before DD, then the capacity is within \SI{1}{bit/s/Hz} of the capacity with coherent detection.
This motivates DD with bipolar modulation (BM) or even complex-valued modulations~\cite{mecozzi_capacity_amdd_2018,SecondiniDirectDetectionBPAM2020,tasbihi_capacity_waveform_channels_time-limited_2020,tasbihi2021direct,KK_receiver_mecozzi_2016,plabst2022achievable}.

As in~\cite{plabst2022achievable}, we study bandlimited systems and compute achievable information rates for frequency-domain raised-cosine (FD-RC) pulses. BM and complex-valued modulations exhibit significant energy gains over classic IM, but the gains in~\cite{plabst2022achievable} are achieved with joint detection and decoding (JDD) which is usually infeasible. The primary goal of this paper is to show how to approach JDD rates with practical detectors.

\subsection{Detection and Decoding}
One can reduce JDD complexity in several ways.
\begin{itemize}
\item Separate detection and decoding (SDD): compute symbol-wise \emph{a posteriori} probabilities (APPs) and pass them to a decoder; see~\cite{sheik_achievable_2017,liga_information_2017}. SDD is usually suboptimal and exhibits a rate loss that grows with the channel memory; see~\cite[Fig.~5]{PfisterAIRFiniteStateChan2001},~\cite[Fig.~7-9]{muller_capacity_separate2004}.
\item Turbo detection and decoding (TDD): use SDD in a turbo loop to exchange extrinsic APPs between the detector and decoder; see~\cite{douillard1995iterative,wang1999iterative}. TDD can approach JDD performance only by carefully designing codes; see~\cite{ten2004design}.
\item Successive interference cancellation (SIC): use SDD with multi-level coding (MLC) and multi-stage detection/decoding (MSD); see~\cite{wachsmann_multilevel_1999,PfisterAIRFiniteStateChan2001,soriaga_determining_2007}. 
\end{itemize}
We focus on SIC because it can approach JDD performance with ``off-the-shelf'' binary codes~\cite{PfisterAIRFiniteStateChan2001,soriaga_determining_2007}.

Besides comparing JDD and SIC performance, we also compare two detection algorithms. First, a standard approach to compute APPs is the forward-backward algorithm (FBA)~\cite{bcjr_1974}.
However, implementing the FBA is infeasible for large memory or symbol alphabets. To reduce complexity, we approximate the APPs by Gibbs sampling, a Markov Chain Monte Carlo method~\cite[Ch.~29]{mackay2003information}. Gibbs sampling has been applied to code-division multiple access~\cite{buchoux2000turbo,wang2000adaptive}, multi-input multi-output systems~\cite{schmidl2000interference,shi2004markov}, and channels with intersymbol interference (ISI)~\cite{yang2001turbo,farhang2006markov,peng2009markov,peng2009low,kashif2008monte}.

\subsection{Organization}
This paper is organized as follows. Sec.~\ref{sec:notation} describes notation, and Sec.~\ref{sec:system-model} reviews the system model. Sec.~\ref{sec:ir} reviews the information rates of SDD and SIC. Sec.~\ref{sec:binary_coding} describes SIC for binary forward error control codes. Sec.~\ref{sec:gibbs_sampling} reviews Gibbs sampling for APP estimation and achievable rates for mismatched decoding. Sec.~\ref{sec:numerical_results} presents numerical results for SDD, SIC, and polar codes and compares them with the plots in~\cite{plabst2022achievable}. The results show that SDD does not approach the JDD rates and that BM rates can be worse than IM rates. The results also show that SIC with Gibbs sampling recovers most gains possible with JDD at low to intermediate signal-to-noise ratio (SNR) but stalls at high SNR. Sec.~\ref{sec:conclusion} concludes the paper and suggests research problems.

\section{Notation}
\label{sec:notation}
Column vectors and matrices are written using bold letters. 
The transpose of the vector $\bm{a}$ is $\bm{a}^{\T}$,
the $n$-dimensional all-zeros vector is $\mnull{n}$, and
the $n \times n$ identity matrix is $\mident{n}$. The element-wise absolute values of the entries in $\bm{a}$ are $\lvert\bm{a}\rvert$ and the element-wise squares are $\bm{a}^{\circ 2}$.
We write strings as $x_\kappa^n=(x_\kappa,\ldots,x_n)$ and $\mathbf{x}_\kappa^n= (\mathbf{x}_\kappa,\ldots,\mathbf{x}_n)$ and omit the subscript if $\kappa = 1$. We write $\mathbf{a}_{[i]}$ for the vector $\mathbf{a}$ without entry $a_i$. For complex-valued $x$, we write $\angle x$ for the phase of $x$.

The sinc function is $\sinc(t) = \sin(\pi t)/(\pi t)$. The signal $a(t)$ and its Fourier transform $A(f)$ are related by $a(t)$ \laplace\, $A(f)$. The expression $g(t)*h(t)$ refers to the convolution of $g(t)$ and $h(t)$ and $\|a(t)\|^2 = \int_{-\infty}^{\infty} \big|\, a(t) \big|^2 \mathrm{d}t $ is the energy of $a(t)$.

Random variables are written with upper-case letters and their realizations with lower-case letters. The probability mass function (PMF) or density of a discrete or continuous random vector $\bm{X}$ is written as $P_{\bm{X}}$ or $p_{\bm{X}}$, respectively. $\E[f(X,Y)]$ is the expectation of $f(X,Y)$. A multivariate real Gaussian density is written as $\GaussDist{\mathbf{x}}{\boldsymbol{\upmu}}{\mathbf{C}}$ where $\boldsymbol{\upmu}$ and $\mathbf{C}$ are a mean vector and covariance matrix, respectively.
The conditional density of $\bm{Y}$ given $\bm{X}$ is written as $p_{\bm{Y}\lvert \bm{X}}$.

The entropy of a discrete-valued $\mathbf{X}$ and the mutual information of $\mathbf{X}$ and $\mathbf{Y}$ with conditional density $p(\mathbf{y}|\mathbf{x})$ are
\begin{align*}
    & H(\bm{X}) = \E \left[-\log_2 P(\bm{X})\right] \\
    & I(\bm{X};\bm{Y}) = \E \left[ \log_2 \frac{p(\mathbf{Y}\lvert \mathbf{X})}{p(\mathbf{Y})}\right]
\end{align*}
where we measure the quantities in bits. As mentioned above, we discard subscripts on PMFs or densities if the arguments are uppercase or lowercase versions of their random variables.
For $n$-dimensional $\bm{X}$, define the entropy rate, conditional entropy rate, and mutual information rate in bits per transmitted symbol as the respective
\begin{align*}
    & H_n(\bm{X}) = \frac{1}{n}  H(\bm{X}), \quad H_n(\bm{X}\lvert \bm{Y}) = \frac{1}{n}  H(\bm{X} \lvert \bm{Y}) \\
    & I_n(\bm{X};\bm{Y}) = H_n(\bm{X}) -  H_n(\bm{X}|\bm{Y}).
\end{align*}

\section{System Model}
\label{sec:system-model}

Propagation in fiber is described by the Nonlinear Schr\"odinger Equation~\cite[p.~65]{AgrawalFourthEdFiberOptics} that models attenuation and chromatic dispersion (CD), both linear effects, and a Kerr nonlinearity. We consider short-reach links without optical amplifiers or optical noise and with sufficiently small launch power so we can neglect the Kerr non-linearity. 

\subsection{Continuous-Time Model}
\label{sec:time-continuous-model}
\begin{figure*}[t!]
    \centering
    \usetikzlibrary{decorations.markings}
\tikzset{node distance=2.5cm}

\pgfdeclarelayer{background}
\pgfdeclarelayer{foreground}
\pgfsetlayers{background,main,foreground}
\tikzset{boxlines/.style = {draw=black!20!white,}}
\tikzset{boxlinesred/.style = {densely dashed,draw=red!50!white,thick}}

\pgfmathsetmacro{\samplerwidth}{30}

\tikzset{midnodes/.style = {midway,above,text width=1.5cm,align=center,yshift=-0.1em}}
\tikzset{midnodesRP/.style = {midway,above,text width=1.5cm,align=center,yshift=-1.4em}}

\begin{tikzpicture}[]
    \footnotesize
    \node[comblock] (diff) {$f_\mathrm{diff}(\cdot)$};
    \node [comblock,right of=diff,node distance=2cm] (txfilter) {$g_\text{tx}(t)$};
    \node [comblock,right of=txfilter,node distance=2.2cm] (cir) {$h_L(t)$};
    \node [comblock,right of=cir,node distance=2.3cm] (sld) {$\left\lvert \,\cdot\, \right\rvert^2$};
    \node [sum,right of=sld,node distance=1.8cm] (sumnode) {$+$};
    \node [input, name=noise,below of=sumnode,node distance=0.5cm] {Input};
    \node [comblock,right of=sumnode,node distance=1.9cm] (rxfilter) {$g_\text{rx}(t)$};
    \node [comblock,right of=rxfilter,minimum width=\samplerwidth pt,node distance=2.1cm] (sampler) {};
    \node [comblock,right of=sampler,node distance=2.2cm] (dsp) {Detector};
    \node [input, name=output, right of=dsp,node distance=1.6cm] {Output};
    \draw[thick] (sampler.west) -- ++(\samplerwidth/4 pt,0) --++(\samplerwidth/2.7 pt,\samplerwidth/4.5 pt );
    \draw[thick] (sampler.east) -- ++(-\samplerwidth/3pt,0);

    \draw ($(sampler.west) + (\samplerwidth/4.5 pt,0.25)$)edge[out=0,in=100,-latex,thick] ($(sampler.east) + (-\samplerwidth/2.5 pt,-0.2)$);

    \draw[latex-,thick] (diff) --++ (-1.2,0) node[midnodes](){$U_\kappa$};
    \draw[-latex,thick] (diff) -- node[midnodes](s_ti){$X_\kappa$  } (txfilter);

    \draw[-latex,thick] (txfilter) -- node[midnodes](){$X(t)$\\[0.2em] } (cir);
    \draw[-latex,thick] (cir) -- node[midnodes](){$X_L(t)$\\[0.2em] } (sld);
    \draw[-latex,thick] (sld) -- node[midnodes](){$Z^\prime(t)$\\[0.2em] } (sumnode);
    \draw[-latex,thick] (sumnode) -- node[midnodes](){$Y'(t)$\\[0.2em] } (rxfilter);
    \draw[-latex,thick] (rxfilter) -- node[midnodes](){$Y(t)$\\[0.2em] } (sampler);
    \draw[-latex,thick] (sampler) --  node[midnodes](){$Y_k$\\ [0.2em] } (dsp)  ;
    \draw[-latex,thick] (dsp) --  node[midnodes,xshift=0cm,align=center](){$\hat{U}_\kappa$\\[0.29em] }(output);

    \draw[-latex,thick] (noise) -- (sumnode);
    \node[below] () at (noise) {$N'(t) \in\, \mathbb{R}$};
    \node[above,yshift=1em] () at (sampler) {Sampler};
    \node[below,yshift=-1.5em] () at (sampler) {Rate $2B$};
    
    \node[below,yshift=-1.5em] () at (txfilter) {$\frac{1+\alpha}{2}[-B,B]$};
    \node[below,yshift=-1.5em] () at (rxfilter) {$[-B,B]$};

    \node[below,yshift=-1.5em] () at (cir) {Channel};
    \node[below,yshift=-1.5em] () at (sld) {SLD};

    \begin{pgfonlayer}{background}
        \draw[boxlinesred] ($( txfilter) + (0,+25pt)$) -- node[yshift=0.5cm,xshift=-0.13cm,font=\scriptsize  ,align=right,]{Electrical $\;$ Optical}  ($( txfilter) + (0,-15pt)$);
    \end{pgfonlayer}
    \begin{pgfonlayer}{background}
        \draw[boxlinesred] ($( sld) + (0,+25pt)$) -- node[yshift=0.5cm, xshift=0.13cm,font=\scriptsize  ,align=right,]{Optical $\;$ Electrical}  ($(sld) + (0,-15pt)$);
    \end{pgfonlayer}

\end{tikzpicture}
    \caption{System model with DD and two-fold oversampling~\cite{plabst2022achievable}.}
    \label{fig:continuous_detailed_system_model}
\end{figure*}
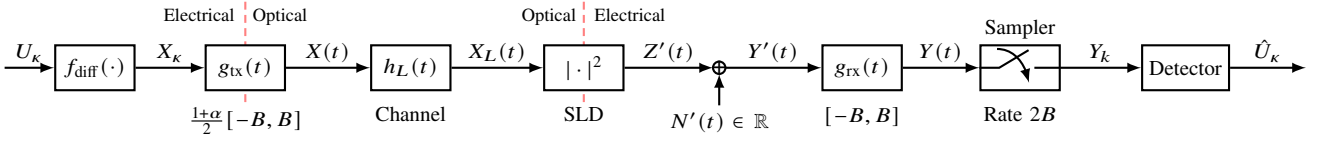

The model is depicted in Fig.~\ref{fig:continuous_detailed_system_model}; see~\cite{plabst2022achievable}.
The symbol alphabet is $\mathcal{A}=\{a_1,\ldots,a_M\}$ with~$M=2^m$.
A source puts out uniformly, independently, and identically distributed (u.i.i.d.) symbols $(U_\kappa)_{\kappa \in \mathbb{Z}}$ with $U_\kappa \in \mathcal{A}$. A differential phase mapper $f_\mathrm{diff}(\cdot)$ maps these symbols to the transmit symbols $(X_\kappa)_{\kappa \in \mathbb{Z}}$, $X_\kappa \in \mathcal{A}$, using $|X_k|=|U_k|$ and $\angle X_k = \angle X_{k-1} + \angle U_k$ for phase modulation. Differential coding helps resolve phase ambiguities; see~\cite[Sec.~IV]{plabst2022achievable} and the Appendix.

For the continuous-time signal, we use an FD-RC pulse shape $g_\text{tx}(t)$ to obtain the baseband waveform 
\begin{align}
    X(t)  = \sum\limits_{\kappa} X_\kappa \cdot g_\text{tx}(t-\kappa T_\text{s})
\end{align}
with symbol-rate $B = 1/T_\text{s}$. The roll-off factor is $\alpha$ and $\alpha=0$ corresponds to the sinc pulse; see~\cite[Eq.~(5)-(6)]{plabst2022achievable}.
The channel exhibits CD with response~\cite[Sec. II.B]{wiener_filter_plabst2020}
\begin{align}
    h_L(t) \;\laplace\; H_L(f) = e^{\mathrm{j} (\beta_2/2) \omega^2 L }
\end{align}
where $\beta_2$ is the group velocity dispersion, $\omega=2\pi f$, and $L$ is the fiber length. The convolution of $X(t)$ and $h_L(t)$ results in~$X_L(t)$.
The receiver PD is modeled as a square-law device (SLD) that outputs the intensity $Z^\prime(t) = \lvert X_L(t) \rvert^2$ and the PD noise $N'(t)$ is a real-valued white Gaussian random process with two-sided power spectral density $N_0/2$.

The signal and noise pass through a bandwidth-limited sampling device, resulting in $Y(t)$. The filter $g_\text{rx}(t)$ has a unit gain frequency response in the interval $[-B,B]$ Hz. The analog-to-digital converter sampling rate is $1/T_\text{s}' = 2B$, i.e., the oversampling factor is $N_\text{os} = T_\text{s}/T_\text{s}'=2$ samples per transmit symbol. Observe that the SLD doubles the bandwidth since squaring in time corresponds to self-convolution in frequency, i.e., the FD-RC bandwidth doubles from $(1+\alpha)B$ to $2(1+\alpha)B$. The filter $g_\mathrm{rx}(t)$ thus removes useful signal components at the band edges unless $\alpha=0$. We will study $\alpha=0$ and $\alpha=0.2$.
 
\subsection{Discrete-Time Model}
\label{sec:time-discrete_model}
Suppose first that $g_\text{tx}(t)$ is a sinc pulse, so the filter $g_\text{rx}(t)$ and two-fold oversampling provide sufficient statistics.
The channel response up to $X_L(t)$ is $\psi(t) = g_\text{tx}(t) * h_L(t)$ and the samples are $\psi_k = \psi(kT_\text{s}'),\, k\in\mathbb{Z}$. For simplicity, suppose $\psi_k=0$ for $k\notin[0,K-1]$, where $K$ is an odd integer. 
We write the channel Toeplitz matrix as $\bm{\Psi} \dimC{2n \times (2n + K-1)}$, which is constructed from the oversampled channel response $\bm{\uppsi} = \left[\psi_{K-1}, \ldots,  \psi_0  \right] \in \mathbb{C}^K$. Consider the vector of noise-free samples
\begin{align}
    \mathbf{z} &= \lvert\mathbf{\Psi} \mathbf{\tilde{x}}' \rvert^{\circ 2} = \left[z_1,\; z_2,\; z_3, \quad \dots \quad z_{2n} \right]^\mathrm{T} \dimR{2n \times 1}
    \label{eq:noise-free_length_n_vector_stacking}
\end{align}
with input $\tilde{\mathbf{x}}'=[\bm{s}_0^{\T}, (\bm{x}')^{\T}]^{\T}$ and upsampled symbols
\begin{align}
    \mathbf{x}' &= \left[0,\;x_1,\; 0,\; x_2,\,\ldots  ,\; 0 ,\; x_{n} \right]^\mathrm{T} \dimC{2n \times 1}
\end{align}
where the initial state vector is
\begin{align}
    \bm{s}_0 = \left[0,\; x_{1-\widetilde{K}},\;0,\; x_{2-\widetilde{K}}, \quad \ldots, \quad 0,\; x_0\right]^{\T} \dimC{(K-1) \times 1}
    \label{eq:initial_state_vector_known}
\end{align}
and the channel memory is $\smash{\widetilde{K}} = (K-1)/2$. 
The discrete-time channel is Gaussian with conditional density
\begin{align}
p(\bm{y}|\bm{x}) = \GaussDist{\bm{y} - \left\lvert \bm{\Psi}
    \tilde{\bm{x}}'
\right\rvert^{\circ 2}} {\mathbf{0}_{2n}}{N_0 B \, \mident{2n}}.
\label{eq:channel}
\end{align}

More generally, for $\alpha \ge 0$ we compute $\bm{\Psi}$ for the continuous-time model described in Sec.~\ref{sec:time-continuous-model}, and we perform simulations with $N_\text{os}=4$ times oversampling as in~\cite[Eq.~(60)]{plabst2022achievable} followed by low-pass filtering and downsampling.

\section{Information Rates}
\label{sec:ir}

We study the information rates of several detection algorithms, including for a fixed block length $n$ and for $n \rightarrow \infty$. The limiting rates are listed in Table~\ref{tab:rates} for convenience.

{\renewcommand{\arraystretch}{1.2}%
\begin{table}
    \centering
    \caption{Information rates studied in the paper.}
    \begin{tabular}{| c || c | c | c |} 
        \hline
                 & Information & Mismatched &  \\
        Detector & Rate        & Rate       & Reference \\
        \hline\hline
        JDD & $I_\text{JDD}$ & $I_\text{$q$,JDD}$ & \eqref{eq:mi_rate_full}, \eqref{eq:mismatch_rate_limit} \\
        \hline
        SDD & $I_\text{SDD}$ & $I_\text{$q$,SDD}$ & \eqref{eq:mi_rate_sapp}, Sec.~\ref{sec:mismatched-decoding} \\ 
        \hline
        SIC / MSD & $I_\text{SIC}=I_\text{MSD}$ & $I_\text{$q$,SDD}=I_\text{$q$,MSD}$ & \eqref{eq:MI_sic_n_infinity}, Sec.~\ref{sec:mismatched-decoding} \\
        \hline
        SIC, bit-wise & $I_\text{b-SIC}$ & $I_\text{$q$,b-SIC}$ & \eqref{eq:bsic_inequality}, Sec.~\ref{sec:mismatched-decoding} \\ 
        \hline
    \end{tabular}
    \label{tab:rates}
\end{table}
}

\subsection{SDD Information Rates}
\label{sec:sdd}
SDD computes the symbol-wise APPs $P(u_\kappa|\mathbf{y})$, $\kappa = 1,\ldots,n$; see~\cite{muller_capacity_separate2004}. Consider the bound
\begin{align}
    I_n(\bm{U};\bm{Y}) &\ge H_n(\bm{U}) - \frac{1}{n}\sum_{\kappa=1}^n H(U_\kappa|\bm{Y}) := I_\text{$n$,SDD}. \label{eq:separate_detection_decoding_rate} 
\end{align}
The JDD and SDD limiting rates are
\begin{align}
    I_\text{JDD} &= \lim_{n \rightarrow \infty }  H_n(\bm{U}) -  H_n(\bm{U}|\bm{Y})
    \label{eq:mi_rate_full}\\
     I_\text{SDD} &= \lim_{n\rightarrow \infty} H_n(\bm{U}) - \frac{1}{n}\sum_{\kappa=1}^n H(U_\kappa|\bm{Y})
    \label{eq:mi_rate_sapp}
\end{align}
and~\eqref{eq:separate_detection_decoding_rate} implies $I_\text{SDD} \le I_\text{JDD}$. The difference $I_\text{JDD}-I_\text{SDD}$ may be significant for channels with memory because SDD puts out marginals which neglect dependencies; see~\cite[Figs.~7-9]{muller_capacity_separate2004},~\cite[Fig.~12]{kavcic_binary_2003}.

\subsection{SIC Information Rates}
\label{sec:sic}
\begin{figure*}
    \centering
    \begin{subfigure}[t]{0.5\textwidth}
    \centering
    \usetikzlibrary{decorations.markings}
\tikzset{node distance=0.9cm}

\pgfdeclarelayer{background}
\pgfdeclarelayer{foreground}
\pgfsetlayers{background,main,foreground}

\pgfmathsetmacro{\samplerwidth}{30}

\tikzset{redbox/.style = {rounded corners=2pt,draw,font=\footnotesize,minimum height=0.9cm,minimum width=0.9cm,fill=mycolor5bright,}}
\tikzset{bluebox/.style = {rounded corners=2pt,draw,font=\footnotesize,minimum height=0.9cm,minimum width=0.9cm,fill=mycolor1bright,}}
\tikzset{bbox/.style = {rounded corners=2pt,draw,font=\footnotesize,minimum height=0.9cm,minimum width=0.9cm,fill=mycolor7bright,}}
\tikzset{greenbox/.style = {rounded corners=2pt,draw,font=\footnotesize,minimum height=0.9cm,minimum width=0.9cm,fill=mycolor6bright}}
\tikzset{dot/.style = {anchor=base,fill,circle,inner sep=1pt}}

\begin{tikzpicture}[scale=0.9, transform shape]
\renewcommand{\baselinestretch}{1}
\footnotesize

\node[redbox] (u1) {$U_1$};
\node[bluebox,below of=u1] (u2) {$U_2$};
\node[bbox,below of=u2] (u3) {$U_3$};
\node[greenbox,below of=u3] (uM) {$U_4$};

\node[redbox,right of=u1] (uM1) {$U_5$};
\node[bluebox,below of=uM1] (uM2) {$U_6$};
\node[bbox,below of=uM2] (uM3) {$U_7$};
\node[greenbox, below of=uM3] (u2M) {$U_8$};

\node[right of=uM1,redbox] (u2M1) {$U_9$};
\node[below of=u2M1,bluebox] (u2M2) {$U_{10}$};
\node[bbox,below of=u2M2] (u2M3) {$U_{11}$};
\node[greenbox,below of=u2M3] (u3M) {$U_{12}$};

\node[redbox,right of=u2M1,minimum width=0.9cm] (u3M1) {$U_{13}$};
\node[bluebox,below of=u3M1,minimum width=0.9cm] (u3M2) {$U_{14}$};
\node[bbox,below of=u3M2] (u3M3) {$U_{15}$};
\node[greenbox,below of=u3M3,minimum width=0.9cm] {$U_{16}$};

\node[redbox,right of=u3M1,minimum width=0.9cm] (u4M1) {$U_{17}$};
\node[bluebox,below of=u4M1,minimum width=0.9cm] (u4M2) {$U_{18}$};
\node[bbox,below of=u4M2] (u4M3) {$U_{19}$};
\node[greenbox,below of=u4M3,minimum width=0.9cm] {$U_{20}$};

\draw[thick,red,line width=1.3pt,rounded corners=2pt] ($(u3) + (-0.4,+0.4)$) -- ($(uM3) + (0.4,0.4)$) -- ($(uM3) + (0.4,-0.4)$) --  ($(u3) + (-0.4,-0.4)$) -- cycle ; 

\draw[thick,black,densely dashed,line width=1.3pt,rounded corners=2pt ] ($(u2M3) + (-0.4,+0.4)$) -- ($(u2M3) + (0.4,0.4)$) -- ($(u2M3) + (0.4,-0.4)$) --  ($(u2M3) + (-0.4,-0.4)$) -- cycle ; 

\draw[thick ] (u1) -- (-0.8,+0.8) node[above left]{$V_{s,\sicindex}$};
\draw[thick ] ($(u1) + (-0.45,+0.45)$) -- node[above](m){$s$} (-1.5,+0.45);
\draw[thick ] ($(u1) + (-0.45,+0.45)$) -- node[left](r){$\sicindex$} (-0.45,+1.2);

\draw[-latex] ($(m) + (0,-1.5)$) -- ($(m) + (0,-2)$)  node[left,rotate=90]{Stage}; 
\draw[-latex] ($(r) + (+1.0,0)$) -- ($(r) + (2,0)$) node[right]{Within stage};

\end{tikzpicture}
    \caption{S/P conversion of an input string $U^{20}$ to $V_{s,\sicindex}$, $s = 1,\ldots, 4$ and $\sicindex=1,\ldots,5$.}   
    \label{fig:SP_conversion}
    \end{subfigure}%
    \begin{subfigure}[t]{0.5\textwidth}
    \centering
    \begin{tikzpicture}[]
\renewcommand{\baselinestretch}{1}

\begin{axis}[%
width=1\textwidth,
height=0.6\textwidth,
scale only axis,
xmin=-1.2,
xmax=5,
tick align=outside,
ymin=0,
ymax=6.5,
zmin=0,
zmax=1,
view={-106.219384615385}{45},
axis background/.style={fill=white},
axis x line*=bottom,
axis y line*=left,
axis z line*=left,
xmajorgrids,
ymajorgrids,
zmajorgrids,
legend style={at={(1.03,1)}, anchor=north west, legend cell align=left, align=left, draw=white!15!black,
axis lines=none,
}
]

\definecolor{mybordercolor}{HTML}{6E605E}%
\definecolor{stemcolor}{HTML}{3D3D3D}%

\filldraw[
        rounded corners=2pt,
        draw=mybordercolor,
        fill=mycolor5bright,%
    ]          (3.5,  0.5, 0)
            -- (4.5,  0.5, 0)
            -- (4.5,  5.5, 0)
            -- (3.5,  5.5, 0)
            -- cycle;
        
\filldraw[
        rounded corners=2pt,
        draw=mybordercolor,
        fill=mycolor1bright,%
    ]          (2.5,  0.5, 0)
            -- (3.5,  0.5, 0)
            -- (3.5,  5.5, 0)
            -- (2.5,  5.5, 0)
            -- cycle;

\filldraw[
        rounded corners=2pt,
        draw=mybordercolor,
        fill=mycolor7bright,%
    ]          (1.5,  0.5, 0)
            -- (2.5,  0.5, 0)
            -- (2.5,  5.5, 0)
            -- (1.5,  5.5, 0)
            -- cycle;

\filldraw[
    rounded corners=2pt,
        draw=mybordercolor,
        fill=mycolor6bright,%
    ]          (0.5,  0.5, 0)
            -- (1.5,  0.5, 0)
            -- (1.5,  5.5, 0)
            -- (0.5,  5.5, 0)
            -- cycle;

\draw[solid,mybordercolor] (0.5,4.5,0) --  (4.5,4.5,0); %
\draw[solid,mybordercolor] (0.5,3.5,0) --  (4.5,3.5,0); %
\draw[solid,mybordercolor] (0.5,2.5,0) --  (4.5,2.5,0); %
\draw[solid,mybordercolor] (0.5,1.5,0) --  (4.5,1.5,0); %

\draw[ultra thick,solid,red,rounded corners=2pt,] (1.5,5.5,0) --  (2.5,5.5,0) -- (2.5,3.5,0) -- (1.5,3.5,0) -- cycle; 

\draw[black,densely dashed,ultra thick,rounded corners=2pt,] (1.5,3.5,0) --  (2.5,3.5,0) --  (2.5,2.5,0) --  (1.5,2.5,0) --cycle; %

\addplot3 [ ycomb, shadowed, color=stemcolor, forget plot,line width=0.8pt, mark=*, mark options={ solid, mark size=1.3pt, stemcolor}]
 table[row sep=crcr] {%
1	1	0.0116820913936517\\
2	1	0.0056640414682982\\
3	1	0.00788919655231655\\
4	1	0.0111089619055566\\
1	2	0.0161021580430195\\
2	2	0.0259694191541276\\
3	2	0.0484354806189115\\
4	2	0.114996815909359\\
1	3	0.34830629268136\\
2	3	0.856712052039897\\
3	3	0.34830629268136\\
4	3	0.114996815909359\\
1	4	0.0484354806189115\\
2	4	0.0259694191541277\\
3	4	0.0161021580430195\\
4	4	0.0111089619055566\\
1	5	0.00788919655231658\\
2	5	0.00566404146829827\\
3	5	0.0116820913936517\\
4	5	0.0286419170014503\\
};

\draw[-latex] (5.1,5.8,0) -- (5.1,4.9,0) node[font=\footnotesize,above ,xshift=0.1cm,yshift=-0.05cm,rotate=5]{Within Stage }; 
\draw[-latex] (5.1,5.8,0) -- (3.4,5.8,0) node[font=\footnotesize,rotate=-46,yshift=0.05cm,xshift=-0.2cm,below]{Stage};

\node[] at(2,3,0.98671205203989) {$\lvert \psi_\kappa \rvert$}; 

\draw[thick,-] (1.5,2.7,0) -- (-0.3,2.3,0) node[right]{$V_{3,3} = U_{11}$};

\end{axis}
\end{tikzpicture}%
    \caption{ISI with SIC.}   
    \label{fig:SIC_memory_decoupling}
    \end{subfigure}
    \caption{SIC example.}
    \label{fig:SIC_left_right}
\end{figure*}
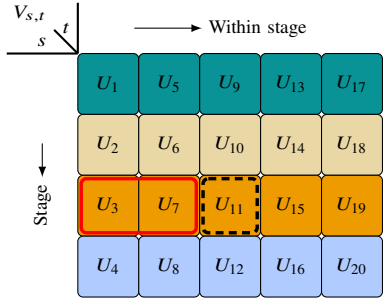
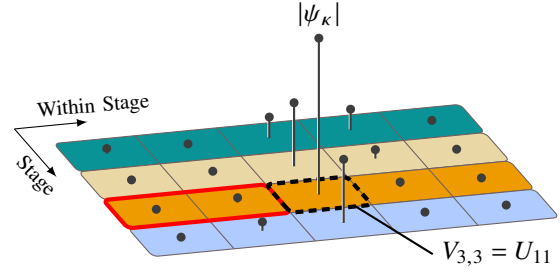

SIC performs a serial-to-parallel (S/P) conversion on the information $\bm{U}$, i.e., downsample $\bm{U}$ to create $S$ shorter strings of length $\siclength = n/S$ (assume $\siclength \in \mathbb{Z}$). We write
\begin{align}
    \bm{V}_{s}
    = \big( V_{s,\sicindex} \big)_{\sicindex=1}^{\siclength}
    = \big(U_{s}, U_{s+S}, \ldots, U_{s+(N-1)S}   \big) 
    \label{eq:subsampling}
\end{align}
and form the vector $\bm{V} = ( \bm{V}_s )_{s=1}^S$. Fig.~\ref{fig:SP_conversion} illustrates the S/P conversion for an input string $U^{20}$ with $S=4$ and $\siclength=5$. The four rows in Fig.~\ref{fig:SP_conversion} represent $\mathbf{V}_1,\ldots,\mathbf{V}_4$. 

We relate the JDD, SDD, and SIC rates; see~\cite{soriaga_determining_2007,PfisterAIRFiniteStateChan2001}. We have $I_n(\bm{U};\bm{Y}) =I_n(\bm{V};\bm{Y})$ and the bounds
\begin{align}
    I_n(\bm{V};\bm{Y}) &\stackrel{(a)}{=} H_n(\bm{V}) - \frac{1}{n}\sum_{s=1}^S H(\bm{V}_s|\bm{Y},\bm{V}^{s-1}) 
    \label{eq:mi_full_rate_n}
    \\
    &\stackrel{(b)}{\geq} \underbrace{H_n(\bm{V}) - \frac{1}{n}\sum_{s=1}^S \sum_{\sicindex=1}^{\siclength} H(V_{s,t}|\bm{Y},\bm{V}^{s-1})}_{\displaystyle :=\, I_\text{$n$,SIC}} \label{eq:sic_inequality_alt}\\
    &\stackrel{(c)}{\geq} \underbrace{H_n(\bm{V}) - \frac{1}{n}\sum_{s=1}^S \sum_{\sicindex=1}^{\siclength} H( V_{s,\sicindex} | \bm{Y} )}_{\displaystyle =\, I_\text{$n$,SDD}} \label{eq:sapp_inequality} 
\end{align}
where step $(a)$ follows from the chain rule of entropy; steps $(b)$-$(c)$ follow because conditioning cannot increase entropy; and step $(c)$ recovers the SDD rate. The limiting SIC rate is
\begin{align}
    I_\text{SIC} = \lim_{n\rightarrow \infty} I_\text{$n$,SIC}.
    \label{eq:MI_sic_n_infinity}
\end{align}

Fig.~\ref{fig:SIC_memory_decoupling} shows how SIC reduces interference. The colored strips are the same as in Fig.~\ref{fig:SP_conversion} and the points show the magnitudes $|\psi_k|$ of the channel response samples due to the symbol $V_{3,3}=U_{11}$.
SIC successively detects and decodes the rows in Fig.~\ref{fig:SP_conversion} with the vectors $\mathbf{V}_1,\ldots,\mathbf{V}_4$. The rates~\eqref{eq:mi_full_rate_n} and~\eqref{eq:sic_inequality_alt} are the same if $V_{s,1},\dots,V_{s,N}$ are independent given $\mathbf{Y}$ and $\mathbf{V}^{s-1}$ for each $s$.
For example, consider $\mathbf{V}_3$. The channel outputs $\mathbf{Y}$ that are relevant for detecting $V_{3,3}=U_{11}$ are hardly affected by $V_{3,1}=U_3$ and $V_{3,2}=U_7$.

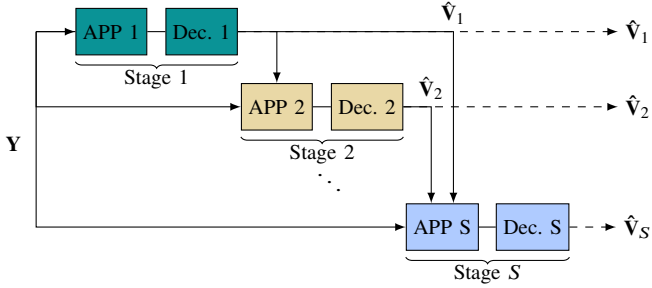
\begin{figure}[!t]
    \centering
    \usetikzlibrary{decorations.markings}
\tikzset{node distance=1cm}

\pgfdeclarelayer{background}
\pgfdeclarelayer{foreground}
\pgfsetlayers{background,main,foreground}

\pgfmathsetmacro{\samplerwidth}{30}

\tikzset{redbox/.style = {draw,minimum height=2.2em,minimum width=3.2em,fill=mycolor5bright}}
\tikzset{bluebox/.style = {draw,minimum height=2.2em,minimum width=3.2em,fill=mycolor1bright}}
\tikzset{greenbox/.style = {draw,minimum height=2.2em,minimum width=3.2em,fill=mycolor6bright}}
\tikzset{dot/.style = {anchor=base,fill,circle,inner sep=1pt}}
\tikzstyle{point}=[fill,shape=circle,minimum size=3pt,inner sep=0pt]

\begin{tikzpicture}
\renewcommand{\baselinestretch}{1}
\footnotesize

\node[] (y) {};
\node[redbox,right of=y,align=center,node distance=1.0cm,font=\footnotesize] (app1) {APP 1};
\node[redbox,right of=app1,align=center,node distance=1.2cm] (dec1) {Dec. 1};
\node[bluebox,below of=app1,xshift=2.2cm,align=center] (app2) {APP 2};
\node[bluebox,right of=app2,node distance=1.2cm,align=center] (dec2) {Dec. 2};

\node[below of=app2,node distance=0.9cm,xshift=0.7cm] (recDots) {$\ddots$};
\node[greenbox,below of=recDots,node distance=0.7cm,xshift=1.5cm,align=center] (appM) {APP S};
\node[greenbox,right of=appM,node distance=1.2cm,align=center] (decM) {Dec. S};

\draw (app1 -|  y) -- (app1) --(dec1) node[midway] (mid) {};

\draw[-latex] (app1 -| y) |- (appM);
\draw[-latex] (app1 -| y) |- (app2);

\node[xshift=-0.3cm,yshift=-0.5cm](y_desc) at(y |- app2) {$\mathbf{Y}$}; 

\draw[-latex] (dec1) -| (appM.65) node[midway,above] {$\hat{\mathbf{V}}_1$};
\draw[-latex] (app2 |- app1) -- (app2);
\draw[-latex] (app2) -- (dec2) -| (appM.115) node[midway,above] {$\hat{\mathbf{V}}_2$};

\node[xshift=1.4cm,font=\footnotesize](out1) at(dec1 -| decM) {$\mathbf{\hat{V}}_1$}; 
\node[xshift=1.4cm,font=\footnotesize](out2) at(dec2 -| decM) {$\mathbf{\hat{V}}_2$}; ; 
\node[xshift=1.4cm,font=\footnotesize](outM) at(decM -| decM) {$\mathbf{\hat{V}}_S$}; ;

\draw[-latex] (app1 -|  y) -- (app1);
\draw (appM) -- (decM); 
\path[] (app1) -- (appM) node[midway] (mid2) {};

\draw[-latex,dashed] (dec1) -- (out1);
\draw[-latex,dashed] (dec2) -- (out2);
\draw[-latex,dashed] (decM) -- (outM);

\draw [decorate, decoration = {mirror,brace}] ($(app1.west) + (0,-0.4)$) -- node[midway,below,font=\footnotesize] {Stage 1} ($(dec1.east) + (0,-0.4)$);
\draw [decorate, decoration = {mirror,brace}] ($(app2.west) + (0,-0.4)$) -- node[midway,below,font=\footnotesize] {Stage 2} ($(dec2.east) + (0,-0.4)$);
\draw [decorate, decoration = {mirror,brace}] ($(appM.west) + (0,-0.4)$) -- node[midway,below,font=\footnotesize] {Stage $S$} ($(decM.east) + (0,-0.4)$);

\end{tikzpicture} 
    \vspace{-3pt}
    \caption{SIC receiver structure.}   
    \label{fig:decoding_scheme}
\end{figure}

We describe some properties of the SIC rates. First, we write $I_\text{$n$,SIC}$ in~\eqref{eq:sic_inequality_alt} as 
\begin{align}
    I_\text{$n$,SIC}
    = \frac{1}{S} \sum_{s=1}^S  \underbrace{\frac{1}{\siclength} \sum_{\sicindex=1}^{\siclength} I\left(V_{s,\sicindex} ; \bm{Y},\bm{V}^{s-1}\right)}_{\displaystyle := I^{(s)}_\text{$\siclength$,SIC} }. 
    \label{eq:sic_stage_m}
\end{align}
The expression~\eqref{eq:sic_stage_m} suggests the receiver in Fig.~\ref{fig:decoding_scheme} with $S$ stages, each with a symbol-wise APP detector and decoder. 
In SIC stage $s$, the detector computes APPs 
\begin{align}
    P(v_{s,\sicindex}\lvert \mathbf{y}, \mathbf{\hat{v}}^{s-1}), \quad \sicindex = 1,\ldots, \siclength
    \label{eq:symbol-wise_apps_conditional}
\end{align}
where $\mathbf{\hat{V}}^{s-1}$ is the vector of decoded symbols from previous stages.
Next, suppose $\mathbf{V}_s$ is encoded with a rate less than the limiting rate~\cite[Sec. III a)]{PfisterAIRFiniteStateChan2001}
\begin{align}
 I^{(s)}_\text{SIC} = \lim_{\siclength\rightarrow \infty} I^{(s)}_\text{$\siclength$,SIC}. \label{eq:sic_rate_per_stage}
\end{align}
This allows reliable decoding as the block length grows, and we may assume $\mathbf{V}_s = \mathbf{\hat{V}}_s$ and the rates of the individual stages are non-decreasing
\begin{align}
    I^{(1)}_\text{SIC} \le I^{(2)}_\text{SIC} \le \ldots I^{(S)}_\text{SIC}. \label{eq:sic_inequality_stages}
\end{align}
By~\eqref{eq:mi_full_rate_n}--\eqref{eq:sapp_inequality} we have  $I_\text{SDD} \le I_\text{SIC} \le I_\text{JDD}$.

We remark that the SIC component codes are shorter than those of SDD (alternatively, the decoding latency is larger if one uses the same code length per stage as for SDD).
Moreover, to prevent error propagation, one must back off from the rate~\eqref{eq:sic_rate_per_stage} as the block length decreases, which reduces the SIC gain.
To mitigate this effect, we use polar codes with successive cancellation list decoding (SCL)~\cite{tal_list_2015} and pass the list across decoding stages~\cite{prinz_successive_2018,karakchieva_joint_2019}; see Sec.~\ref{sec:polar}.

\section{Binary Codes for SIC}
\label{sec:binary_coding}
This section combines SIC with binary codes. Each symbol $V_\kappa$ of $\bm{V}$ is labeled by $m$ bits $b(V_\kappa)$ and we write the string of $mn$ bits of $\bm{V}$ as
\begin{align}
    \mathbf{B} = b(\mathbf{V}) = \left(b(V_1),b(V_2),\ldots,b(V_n)\right).
    \label{eq:Bvector}
\end{align}
The $l^\text{th}$ bit of $V_\kappa$ is $b_l(V_\kappa)$ and the string of $l^\text{th}$ bits of $\mathbf{V}$ is $b_l(\bm{V})=(b_l(V_1),\ldots,b_l(V_n))$. We write
\begin{align}
    b_l^i(\mathbf{V}) =  \left( b_l(\mathbf{V}),\ldots, b_i(\mathbf{V})\right)
\end{align}
for $i\geq l$ when referring to more than one bit-level.

\subsection{SIC Transmitter}\label{sec:sic-encoder}
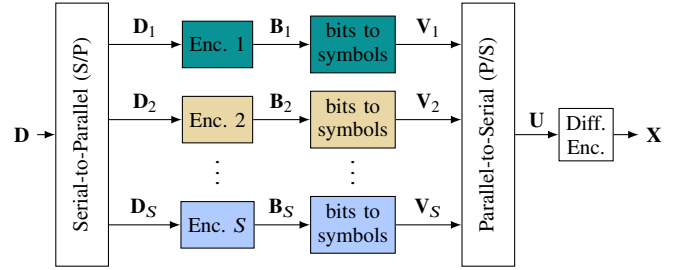
\begin{figure}[t]
    \centering
    \usetikzlibrary{decorations.markings}
\tikzset{node distance=1cm}

\pgfdeclarelayer{background}
\pgfdeclarelayer{foreground}
\pgfsetlayers{background,main,foreground}

\pgfmathsetmacro{\samplerwidth}{30}

\tikzset{redbox/.style = {draw,minimum height=2.2em,minimum width=3.2em,fill=mycolor5bright}}
\tikzset{bluebox/.style = {draw,minimum height=2.2em,minimum width=3.2em,fill=mycolor1bright}}
\tikzset{greenbox/.style = {draw,minimum height=2.2em,minimum width=3.2em,fill=mycolor6bright}}
\tikzset{dot/.style = {anchor=base,fill,circle,inner sep=1pt}}

\begin{tikzpicture}
\renewcommand{\baselinestretch}{1}
\footnotesize
\node[redbox] (Enc1) {Enc. $1$};
\node[bluebox,below of=Enc1] (Enc2) {Enc. $2$};
\node[below of=Enc2,node distance=0.6cm] (EncPoints) {$\vdots$};
\node[greenbox,below of=EncPoints,node distance=0.8cm] (EncM) {Enc. $S$};

\node[redbox,right of=Enc1,node distance=1.8cm,align=center] (Map1) {bits to\\symbols};
\node[bluebox,below of=Map1,align=center] (Map2) {bits to\\symbols};
\node[below of=Map2,node distance=0.6cm] (MapPoints) {$\vdots$};
\node[greenbox,below of=MapPoints,align=center,node distance=0.8cm] (MapM) {bits to\\symbols};

\path[] (Enc1) -- (EncM) node[midway] (mid) {};
\node[draw,minimum height=3.5cm,left of=mid,align=center,node distance=1.8cm,minimum width=0.7cm] (demux) {\rotatebox{90}{Serial-to-Parallel (S/P)}};
\node[draw,minimum height=3.5cm,right of=mid,align=center,node distance=3.6cm,minimum width=0.7cm] (mux) {\rotatebox{90}{Parallel-to-Serial (P/S)}};
\node[left of=demux,node distance=0.8cm] (data) {$\mathbf{D}$};

\node[right of=mux,draw,align=center,node distance=1.3cm] (diff) {Diff.\\Enc.};

\draw[-latex] (data) -- (demux);

\draw[-latex] (demux.east |- Enc1) -- (Enc1) node[midway,above] {$\mathbf{D}_1$};
\draw[-latex] (demux.east |- Enc2) -- (Enc2) node[midway,above] {$\mathbf{D}_2$};
\draw[-latex] (demux.east |- EncM) -- (EncM) node[midway,above] {$\mathbf{D}_S$};

\draw[-latex] (Enc1) -- (Map1) node[midway,above] {$\mathbf{B}_1$};
\draw[-latex] (Enc2) -- (Map2) node[midway,above] {$\mathbf{B}_2$};
\draw[-latex] (EncM) -- (MapM) node[midway,above] {$\mathbf{B}_S$};

\draw[-latex] (Map1) -- (mux.west |- Map1) node[midway,above] {$\mathbf{V}_1$};
\draw[-latex] (Map2) -- (mux.west |- Map2) node[midway,above] {$\mathbf{V}_2$};
\draw[-latex] (MapM) -- (mux.west |- MapM) node[midway,above] {$\mathbf{V}_S$};

\draw[-latex] (mux) -- (diff) node[midway,above] {$\mathbf{U}$};
\draw[-latex] (diff) --++ (0.7,0) node[right] {$\mathbf{X}$};

\end{tikzpicture} 
    \vspace{5pt}
    \caption{SIC encoder for binary codes.}   
    \label{fig:encoding_scheme}
\end{figure}

Fig.~\ref{fig:encoding_scheme} shows the encoder.
A binary source provides $k\leq n \cdot m$ u.i.i.d. bits $\mathbf{D}$. The S/P converter partitions $\mathbf{D}$ into $S$ strings $\mathbf{D}_1, \ldots,\mathbf{D}_S$ where $\mathbf{D}_s$ has length $k_s$. 
Each $\bm{D}_s$ is encoded to a length $\siclength m$ bit string $\bm{B}_s$, so the code rate per stage is $k_s/(\siclength m )$. The bits $\bm{B}_s$ are mapped to the length $\siclength$ string $\bm{V}_s$ of symbols from the modulation alphabet $\mathcal{A}$.
The information rate in bits per channel use (bpcu) for SIC stage $s$ is
\begin{align}
    R_s = \left. k_s \right/ \siclength \text{ [bpcu]} %
    \label{eq:sic_information_rate}
\end{align}
and the overall information rate is
\begin{align}
    R_\text{SIC} = \frac{1}{S} \sum_{s=1}^S R_s = \frac{k}{n}   \text{ [bpcu]}.
\end{align}
Finally, the strings $\mathbf{V}_1,\ldots, \mathbf{V}_S$ are passed to a P/S converter that reverses~\eqref{eq:subsampling} and produces $\mathbf{U}$. 

\subsection{SIC Receiver}
Binary decoders convert symbol-wise APPs to bit-wise APPs. Two common approaches are bit-metric decoding~\cite{martinez2009bit} and MLC/MSD \cite{wachsmann_multilevel_1999}; we focus on the latter. 
Replacing $V_{s,\sicindex}$ by $b(V_{s,\sicindex})$, we write~\eqref{eq:sic_stage_m} as
\begin{align}
    & I_\text{$n$,SIC} =  I_\text{$n$,MSD} \nonumber \\
    & = \frac{1}{S}\sum_{s,l}    \underbrace{ \frac{1}{\siclength} \sum_t 
    I\left(b_l(V_{s,\sicindex});\bm{Y},\bm{V}^{s-1},b_1^{l-1}(V_{s,\sicindex})\right)}_{\displaystyle := I^{(s,l)}_\text{$n$,MSD} } \label{eq:sic_mi_bits_finite_n}
\end{align}
where $I^{(s,l)}_\text{$n$,MSD}$ is the information rate of bit level $l$ in SIC stage $s$. 
We have $m$ binary MSD decoding stages for every SIC stage and work successively through the bit levels.
Binary decoding stage $l$ computes the conditional APP of bit level $l$:
\begin{align}
    P\big(b_l(v_{s,\sicindex})|\bm{y},
    \bm{\hat{v}}^{s-1},\hat{b}_1^{l-1}(v_{s,\sicindex})
    \big), \quad l=1,\ldots,m
    \label{eq:msd-apps}
\end{align}
where $\hat{b}_1^{l-1}(V_{s,\sicindex})$ are the bit estimates from the previous $l-1$ bit levels of $V_{s,\sicindex}$, and given the symbol estimates $\bm{\hat{V}}^{s-1}$ from previous SIC stages. The APPs are passed to a decoder, and the resulting maximum a posteriori probability (MAP) bit estimate $\hat{b}_l(V_{s,\sicindex})$ is passed to the next decoding stage. 

Encoding the bit levels $b_l(V_{s,\sicindex})$ with rate less than
\begin{align}
    I^{(s,l)}_\text{MSD} = \lim_{n\rightarrow \infty} I^{(s,l)}_\text{$n$,MSD}
\end{align}
allows reliable decoding of $b_l(V_{s,\sicindex})$ as the block length grows. By successively decoding over the bit-levels $l = 1,\ldots,m$ and SIC stages $s= 1,\ldots,S$, we assume that $\hat{b}_l(V_{s,\sicindex}) = {b}_l(V_{s,\sicindex})$ and $\bm{\hat{V}}^{s-1} = \bm{V}^{s-1}$. MSD thus achieves the rate $I_\text{SIC}$ in~\eqref{eq:MI_sic_n_infinity}.

\section{Gibbs Sampling Detector}
\label{sec:gibbs_sampling}
The FBA computes the $P(v_{s,\sicindex} \lvert \mathbf{y},\mathbf{\hat{v}}^{s-1})$ and $P(v_{s,\sicindex} \lvert \mathbf{y})$ for~\eqref{eq:sic_inequality_alt} and~\eqref{eq:sapp_inequality}, respectively, and the complexity is exponential in the channel memory $\widetilde{K}$. Gibbs sampling reduces the complexity to quadratic in $\widetilde{K}$, see Table~\ref{tab:complexity}, by generating samples from the joint PMFs $P(\mathbf{v}_s\lvert \mathbf{y}, \mathbf{\hat{v}}^{s-1})$ and $P(\mathbf{v}\lvert \mathbf{y})$.
The $P(v_{s,\sicindex} \lvert \mathbf{y},\mathbf{\hat{v}}^{s-1})$ and $P(v_{s,\sicindex} \lvert \mathbf{y})$  can then be estimated accurately through many such samples. Likewise, the bit-wise APPs~\eqref{eq:msd-apps} for SIC with MSD may be computed using bit-wise sampling~\cite{zhu2005performance,mao2007markov}.

{\renewcommand{\arraystretch}{1.4}%
\begin{table}
    \centering
    \caption{Algorithmic complexity for channel memory $\widetilde K$. The parameter $N_\text{par}$ refers to the number of parallel samplers.}
    \begin{tabular}{|c || c | c |} 
        \hline
        Algorithm & Multiplications per Iteration & Iterations  \\
        \hline\hline
        FBA & $\mathcal{O}\left(  n \cdot S \cdot \lvert \mathcal{A} \rvert^{\widetilde{K}+1}\right)$ & 1 \\
        \hline
        Gibbs (b-SIC) & $\quad\;\mathcal{O}\left(  n \cdot S \cdot m  \cdot \widetilde{K}^2  \cdot N_\text{par} \right)$ & $60$ \\
        \hline
    \end{tabular}
    \label{tab:complexity}
\end{table}
}

\subsection{SIC with Blocks of Bits (b-SIC)}
For SIC, we propose to sample from different APPs than~\eqref{eq:msd-apps}, namely the APPs
\begin{equation}
    P\big(b_l(v_{s,\sicindex})|\bm{y},
    \bm{\hat{v}}^{s-1},\hat{b}_1^{l-1}(\mathbf{v}_{s})
    \big), \quad l=1,\ldots,m. \label{eq:bsic-metric}
\end{equation}
Instead of conditioning only on the first $l-1$ bits of the symbol $v_{s,t}$ as in~\eqref{eq:msd-apps}, we condition on the first $l-1$ bits of all the symbols $\mathbf{v}_s$ from the current stage~$s$. 

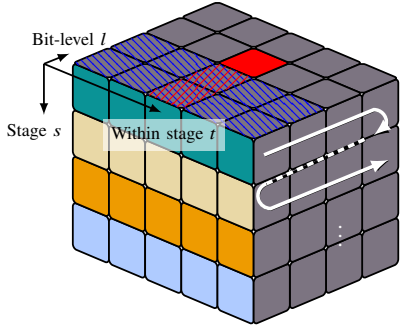
\begin{figure}
    \centering
    \scalebox{0.85}{
    \pgfmathsetmacro\radius{0.1}
  
    \newcommand{\frontcolor}{red}
    \newcommand{\sidecolor}{blue}
            \tdplotsetmaincoords{65}{135}
            \begin{tikzpicture}[scale=0.8]
            \renewcommand{\baselinestretch}{1}
                \clip (-4.9,-3.0) rectangle (4,4.0);
                \begin{scope}[tdplot_main_coords]
                    \foreach \X in {-2.5,-1.5,-0.5,0.5,1.5}{
                        \foreach \Y in {-1.5,-0.5,0.5,1.5}{
                            \ifthenelse{\equal{\Y}{1.5}}{
                            \draw [thick,canvas is yz plane at x=2.5,shift={(\X,\Y)},fill=mycolor5bright] (0.5,0) -- ({1-\radius},0) arc (-90:0:\radius) -- (1,{1-\radius}) arc (0:90:\radius) -- (\radius,1) arc (90:180:\radius) -- (0,\radius) arc (180:270:\radius) -- cycle;
                            }{};
                            \ifthenelse{\equal{\Y}{0.5}}{
                            \draw [thick,canvas is yz plane at x=2.5,shift={(\X,\Y)},fill=mycolor1bright] (0.5,0) -- ({1-\radius},0) arc (-90:0:\radius) -- (1,{1-\radius}) arc (0:90:\radius) -- (\radius,1) arc (90:180:\radius) -- (0,\radius) arc (180:270:\radius) -- cycle;
                            }{};
                            \ifthenelse{\equal{\Y}{-0.5}}{
                            \draw [thick,canvas is yz plane at x=2.5,shift={(\X,\Y)},fill=mycolor7bright] (0.5,0) -- ({1-\radius},0) arc (-90:0:\radius) -- (1,{1-\radius}) arc (0:90:\radius) -- (\radius,1) arc (90:180:\radius) -- (0,\radius) arc (180:270:\radius) -- cycle;
                            }{};
                            \ifthenelse{\equal{\Y}{-1.5}}{
                            \draw [thick,canvas is yz plane at x=2.5,shift={(\X,\Y)},fill=mycolor6bright] (0.5,0) -- ({1-\radius},0) arc (-90:0:\radius) -- (1,{1-\radius}) arc (0:90:\radius) -- (\radius,1) arc (90:180:\radius) -- (0,\radius) arc (180:270:\radius) -- cycle;
                            }{};
     
                            \ifthenelse{\NOT \equal{\X}{-2.5}}{
                            \draw [thick,canvas is xz plane at y=2.5,shift={(\X,\Y)},fill=mycolorcube] (0.5,0) -- ({1-\radius},0) arc (-90:0:\radius) -- (1,{1-\radius}) arc (0:90:\radius) -- (\radius,1) arc (90:180:\radius) -- (0,\radius) arc (180:270:\radius) -- cycle;
                            }{}
                            \ifthenelse{\NOT \equal{\X}{-0.5}}{
                            \draw [thick,canvas is yx plane at z=2.5,shift={(\X,\Y)},fill=mycolorcube] (0.5,0) -- ({1-\radius},0) arc (-90:0:\radius) -- (1,{1-\radius}) arc (0:90:\radius) -- (\radius,1) arc (90:180:\radius) -- (0,\radius) arc (180:270:\radius) -- cycle;
                            }{}
                        }
                    }
                    \draw [thick,canvas is yx plane at z=2.5,shift={(-0.5,-0.5)},fill=red] (0.5,0) -- ({1-\radius},0) arc (-90:0:\radius) -- (1,{1-\radius}) arc (0:90:\radius) -- (\radius,1) arc (90:180:\radius) -- (0,\radius) arc (180:270:\radius) -- cycle;
                    
                    \draw [thick,canvas is yx plane at z=2.5,shift={(-0.5,0.5)},fill=mycolorcube] (0.5,0) -- ({1-\radius},0) arc (-90:0:\radius) -- (1,{1-\radius}) arc (0:90:\radius) -- (\radius,1) arc (90:180:\radius) -- (0,\radius) arc (180:270:\radius) -- cycle;

                    \draw [thick,canvas is yx plane at z=2.5,shift={(-0.5,1.5)},fill=mycolorcube] (0.5,0) -- ({1-\radius},0) arc (-90:0:\radius) -- (1,{1-\radius}) arc (0:90:\radius) -- (\radius,1) arc (90:180:\radius) -- (0,\radius) arc (180:270:\radius) -- cycle;
                    \draw [thick,canvas is yx plane at z=2.5,shift={(-0.5,-1.5)},fill=mycolorcube] (0.5,0) -- ({1-\radius},0) arc (-90:0:\radius) -- (1,{1-\radius}) arc (0:90:\radius) -- (\radius,1) arc (90:180:\radius) -- (0,\radius) arc (180:270:\radius) -- cycle;

                    \pattern [canvas is yx plane at z=2.5,shift={(-2.5,1.5)},pattern=north west lines,pattern color=blue] (0.5,0) -- ({1-\radius},0) arc (-90:0:\radius) -- (1,{1-\radius}) arc (0:90:\radius) -- (\radius,1) arc (90:180:\radius) -- (0,\radius) arc (180:270:\radius) -- cycle;
                    \pattern [canvas is yx plane at z=2.5,shift={(-2.5,0.5)},pattern=north west lines,pattern color=blue] (0.5,0) -- ({1-\radius},0) arc (-90:0:\radius) -- (1,{1-\radius}) arc (0:90:\radius) -- (\radius,1) arc (90:180:\radius) -- (0,\radius) arc (180:270:\radius) -- cycle;
                    \pattern [canvas is yx plane at z=2.5,shift={(-1.5,0.5)},pattern=north west lines,pattern color=blue] (0.5,0) -- ({1-\radius},0) arc (-90:0:\radius) -- (1,{1-\radius}) arc (0:90:\radius) -- (\radius,1) arc (90:180:\radius) -- (0,\radius) arc (180:270:\radius) -- cycle;
                    \pattern [canvas is yx plane at z=2.5,shift={(-1.5,1.5)},pattern=north west lines,pattern color=blue] (0.5,0) -- ({1-\radius},0) arc (-90:0:\radius) -- (1,{1-\radius}) arc (0:90:\radius) -- (\radius,1) arc (90:180:\radius) -- (0,\radius) arc (180:270:\radius) -- cycle;
                    \pattern [canvas is yx plane at z=2.5,shift={(-0.5,0.5)},pattern=north west lines,pattern color=blue] (0.5,0) -- ({1-\radius},0) arc (-90:0:\radius) -- (1,{1-\radius}) arc (0:90:\radius) -- (\radius,1) arc (90:180:\radius) -- (0,\radius) arc (180:270:\radius) -- cycle;
                    \pattern [canvas is yx plane at z=2.5,shift={(-0.5,1.5)},pattern=north west lines,pattern color=blue] (0.5,0) -- ({1-\radius},0) arc (-90:0:\radius) -- (1,{1-\radius}) arc (0:90:\radius) -- (\radius,1) arc (90:180:\radius) -- (0,\radius) arc (180:270:\radius) -- cycle;
                    \pattern [canvas is yx plane at z=2.5,shift={(0.5,0.5)},pattern=north west lines,pattern color=blue] (0.5,0) -- ({1-\radius},0) arc (-90:0:\radius) -- (1,{1-\radius}) arc (0:90:\radius) -- (\radius,1) arc (90:180:\radius) -- (0,\radius) arc (180:270:\radius) -- cycle;
                    \pattern [canvas is yx plane at z=2.5,shift={(0.5,1.5)},pattern=north west lines,pattern color=blue] (0.5,0) -- ({1-\radius},0) arc (-90:0:\radius) -- (1,{1-\radius}) arc (0:90:\radius) -- (\radius,1) arc (90:180:\radius) -- (0,\radius) arc (180:270:\radius) -- cycle;
                    \pattern [canvas is yx plane at z=2.5,shift={(1.5,0.5)},pattern=north west lines,pattern color=blue] (0.5,0) -- ({1-\radius},0) arc (-90:0:\radius) -- (1,{1-\radius}) arc (0:90:\radius) -- (\radius,1) arc (90:180:\radius) -- (0,\radius) arc (180:270:\radius) -- cycle;
                    \pattern [canvas is yx plane at z=2.5,shift={(1.5,1.5)},pattern=north west lines,pattern color=blue] (0.5,0) -- ({1-\radius},0) arc (-90:0:\radius) -- (1,{1-\radius}) arc (0:90:\radius) -- (\radius,1) arc (90:180:\radius) -- (0,\radius) arc (180:270:\radius) -- cycle;
                    
                    \pattern [canvas is yx plane at z=2.5,shift={(-0.5,1.5)},pattern=north east lines,pattern color=red] (0.5,0) -- ({1-\radius},0) arc (-90:0:\radius) -- (1,{1-\radius}) arc (0:90:\radius) -- (\radius,1) arc (90:180:\radius) -- (0,\radius) arc (180:270:\radius) -- cycle;
                    \pattern [canvas is yx plane at z=2.5,shift={(-0.5,0.5)},pattern=north east lines,pattern color=red] (0.5,0) -- ({1-\radius},0) arc (-90:0:\radius) -- (1,{1-\radius}) arc (0:90:\radius) -- (\radius,1) arc (90:180:\radius) -- (0,\radius) arc (180:270:\radius) -- cycle;
                \end{scope}
                
                \newcommand{\xtrans}{-0.6}
                \newcommand{\ytrans}{4.25}
                \draw[thick, -latex] (-3.5+\xtrans,-2+\ytrans) -- (-3.5+\xtrans,-3.0+\ytrans) node[below,font=\footnotesize,xshift=-0.15cm]{Stage $s$}; %
                \draw[thick,-latex] (-3.5+\xtrans,-2+\ytrans) -- (-1.2+\xtrans,-2.95+\ytrans) node[below left,xshift=1cm,yshift=-0.1cm,font=\footnotesize,fill=white,fill opacity=0.6,text opacity=1](t_arrow){Within stage $\sicindex$}; %
                \draw[thick,-latex] (-3.5+\xtrans,-2+\ytrans) -- (-3+\xtrans,-1.8+\ytrans) node[above ,font=\footnotesize]{Bit-level $l$}; %
                
                \node[] at ($(t_arrow) + (1.8,-0.45)$)  (corner)  {};
                
                \draw[-latex,shadowedb,white,ultra thick] 
                (corner) -- ($(corner) + (2.3,0.95)$) 
                arc
                [
                    start angle=90,
                    end angle=-90,
                    x radius=0.25cm,
                    y radius =0.25cm
                ] node[](endr1){};
                
                \draw[-,shadowedb,densely dashed,white,ultra thick] 
                (endr1) -- ($(endr1) - (2.1,0.95)$); 
                
                \draw[-latex,shadowedb,white,ultra thick]  ($(endr1) - (2.1,0.95)$) 
                arc
                [
                    start angle=+110,
                    end angle=270,
                    x radius=0.25cm,
                    y radius =0.25cm
                ] node[](endr2){}
                -- ($(endr2) + (2.3,0.95)$) ; 
                
                \draw[draw opacity=0] ($(endr1) - (2.1,0.95)$) -- node[thick,midway,white,shadowedb,below,xshift=0.2cm,yshift=-0.5cm]{$\mathbf{\vdots}$} ($(endr2) + (2.3,0.95)$); 
             
   \end{tikzpicture}}
    \caption{Binary SIC with $m = 4$ bit levels.}
    \label{fig:b_SIC}
\end{figure}

Consider the illustration in Fig.~\ref{fig:b_SIC} with $S=4$ SIC stages, $\siclength=5$ symbols per stage, and $m=4$ bit-levels per symbol, e.g., 16-ASK modulation.
Consider the APP of the third bit $b_3(V_{1,3})$, marked in red. From~\eqref{eq:msd-apps}, we must condition on the previous bits $(b_1(V_{1,3}),b_2(V_{1,3}))$ marked in shaded red. According to the new APPs~\eqref{eq:bsic-metric}, we condition on the bits in shaded blue.
The advantage of using~\eqref{eq:bsic-metric} is twofold: the new APPs are easier to estimate with sampling, as discussed in the next section, and the achievable rates improve.

To show the rate improvement, observe that for the APPs~\eqref{eq:bsic-metric}, we have
\begin{align}
&I_n(\bm{V};\bm{Y}) \nonumber
=  \frac{1}{n} \sum_{s,l} I\big( b_l(\bm{V}_s) ; \bm{Y},\bm{V}^{s-1}, b_1^{l-1}(\bm{V}_s) \big) \\
& \ge \underbrace{\frac{1}{S} \sum_{s,l} \frac{1}{\siclength} \sum_t I\big( b_l(V_{s,\sicindex}) ; \bm{Y},\bm{V}^{s-1}, b_1^{l-1}(\bm{V}_s) \big)}_{\displaystyle := I_\text{$n$,b-SIC} \text{ via~\eqref{eq:bsic-metric}}} \label{eq:rate_SIC_m-MSD_Gibbs} \\
& \ge \underbrace{\frac{1}{S} \sum_{s,l} \frac{1}{\siclength} \sum_t I\big( b_l(V_{s,\sicindex}) ; \bm{Y},\bm{V}^{s-1}, b_1^{l-1}(V_{s,t}) \big)}_{\displaystyle := I_\text{$n$,SIC} \text{ via~\eqref{eq:msd-apps}}} \label{eq:rate_SIC_m-MSD_Gibbs_fba}. 
\end{align}
The bound~\eqref{eq:rate_SIC_m-MSD_Gibbs_fba} gives the improved rate
\begin{equation}
I_\text{b-SIC} := \lim_{n \rightarrow \infty} I_\text{$n$,b-SIC}
\ge I_\text{SIC}. \label{eq:bsic_inequality}
\end{equation}
Note that $I_\text{$n$,b-SIC}$ depends on the bit mapping, and one may use the FBA to compute the APPs~\eqref{eq:bsic-metric}.

\subsection{Gibbs Sampling with b-SIC}
Consider SIC stage~$s$ and bit stage~$l$ for $\sicindex=1,\ldots,\siclength$. To approximate the APPs~\eqref{eq:bsic-metric}, we sample from the joint posterior PMF of the bits that are not already decoded:
\begin{align}
    P\left( \gibbsvec_{s,l} \,\big|\, \bm{y}, b(\bm{\hat{v}}^{s-1}),\hat{b}_1^{l-1}(\mathbf{v}_s)\right) \label{eq:gibbs_joint_density}
\end{align} 
where
\begin{align}
    \gibbsvec_{s,l} = \pi \left( b_l^m(\mathbf{v}_s), b(\mathbf{v}_{s+1}^S) \right) \label{eq:permutation}
\end{align}
has a total of $W=nm-(s-1)\siclength m - \siclength(l-1)$ bits, namely the $(m-l+1)N$  bits of bit-levels $l,\ldots,m$ of SIC stage $s$ and the $(S-s)mN$ bits of SIC stages $s+1,\ldots,S$. The permutation~$\pi$ sorts the bits in the order they are transmitted over the channel, and this reflects our choice of sampling order; see~\eqref{eq:gibbs_samping_algorithm} and the footnote below. Empirically, this ordering results in faster convergence than other orderings.

To simplify notation, we assume that $s,l$ are fixed and replace $\gibbsvec_{s,l}$ with $\gibbsvec$. We further assume the conditional values are fixed and define
\begin{align}
    \Phi(\gibbsvec) := P(\gibbsvec \,\big|\, \bm{y}, b(\bm{\hat{v}}^{s-1}),\hat{b}_1^{l-1}(\mathbf{v}_s) ).
\end{align}
At iteration $i=0$, we initialize with a realization $\gibbsvec(0)$ of u.i.i.d. bits. At iteration $i$, $i\geq 1$, the sampler successively draws realizations from the distributions 
\begin{align}
    \gibbsscalar_w(i) \sim  \Phi(\gibbsscalar_w \lvert \gibbsscalar^{w-1}(i), \gibbsscalar_{w+1}^W(i-1)) \label{eq:gibbs_samping_algorithm}
\end{align}
 for $w=1,\ldots,W$. Observe that we use samples from iteration $i$ for $\gibbsscalar^{w-1}(i)$ and samples from iteration $i-1$ for $\gibbsscalar_{w+1}^W(i-1)$ when sampling $\gibbsscalar_w(i)$.\footnote{This sampling structure is the reason for choosing the permutation $\pi$ in~\eqref{eq:permutation}. The samples of $\gibbsscalar^{w-1}(i), \gibbsscalar_{w+1}^W(i-1))$ are ordered as the bits affect the channel.}
 The algorithm stops after $N_\mathrm{iter}$ iterations; for $N_\text{iter} \rightarrow \infty$ the distribution of $\mathbf{\gibbsvec}(N_\text{iter})$ tends to $\Phi(\gibbsvec)$~\cite[Ch. 29.5]{mackay2003information}. 

We remark that the expression in~\eqref{eq:gibbs_samping_algorithm} is
\begin{equation}
\begin{aligned}
    & \Phi\big(\gibbsscalar_w \lvert  \gibbsscalar^{w-1}(i), \gibbsscalar_{w+1}^W(i\!-\!1) \big) 
    = \frac{1}{A_1} p\left(\mathbf{y} \lvert \,\mathbf{x}(\gibbsscalar_w)\,    \right)
    \label{eq:gibbs_conditional_probability}
\end{aligned}
\end{equation}
with differentially-encoded symbols
\begin{align}
    \mathbf{x}(\Gibbsscalar_w) = f_\text{diff}\big(b(\mathbf{\hat{V}}^{s-1}), \hat{b}_1^{l-1}(\mathbf{V}_s) \gibbsscalar^{w-1}(i),\Gibbsscalar_w, \gibbsscalar_{w+1}^W(i\!-\!1)\big)
\end{align}
that depend on $\gibbsscalar_w$, and where the normalization constant $A_1$ ensures that~\eqref{eq:gibbs_conditional_probability} is a PMF. One may readily sample from~\eqref{eq:gibbs_conditional_probability} because the channel~\eqref{eq:channel} is Gaussian.
When using $f_\text{diff}(\cdot)$ with binary inputs, we place a bit-to-symbol mapping before the differential phase coding.  

Finally, we explain the complexity advantage of~\eqref{eq:bsic-metric} over~\eqref{eq:msd-apps}. The modified sampling algorithm runs $N_\mathrm{iter}$ iterations through $\gibbsvec_{s,l}$ for $s=1,\ldots,S$ and $l=1,\ldots,m$, i.e., it runs $Sm$ times. This is because the conditioning in~\eqref{eq:bsic-metric} is independent of the index $t$, and we can use the same joint PMF~\eqref{eq:gibbs_joint_density} for each $\sicindex$. Instead, the conditioning of the APPs~\eqref{eq:msd-apps} changes for each $\sicindex$, and we must sample from $N$ joint PMFs, which requires $SmN$ runs of the sampling algorithm.

\subsection{APP Estimation and Stalling}
The bit-wise APPs may be written as
\begin{align}
    \Phi(\gibbsscalar_w) = \sum_{\gibbsvec_{[w]}} \Phi(\gibbsvec)
    = \E \left[ \Phi(\gibbsscalar_w \lvert \Gibbsvec_{[w]})  \right].
    \label{eq:app_pi}
\end{align}
One can approximate the expectation
from the sampling strings $\gibbsvec(0), \ldots, \gibbsvec(N_\text{iter})$ via empirical PMF estimation or importance sampling~\cite{farhang2006markov}. A simple estimate is
\begin{align}
    \Phi(\gibbsscalar_w) \approx \frac{1}{N_\text{iter}} \sum_{i = 1}^{N_\text{iter}} \; \Phi(\gibbsscalar_w \lvert \gibbsvec_{[w]}(i)).
    \label{eq:app_est_simple_monte_c}
\end{align}

The estimate~\eqref{eq:app_est_simple_monte_c} is accurate at low SNR, and decoding achieves near-MAP performance~\cite{farhang2006markov}.
However, the estimate can be inaccurate at high SNR if the sampling stalls because the conditionals~\eqref{eq:gibbs_samping_algorithm} are overconfident, e.g., due to poor initialization or a bad current state of the Gibbs sampler~\cite{farhang2006markov,senst2011rao,hassibi2014optimized,hedstrom_achieving_2017}. For instance, consider $\Phi(\gibbsscalar_w = 0) \approx 1$ and \mbox{$\Phi(\gibbsscalar_w = 1) \approx 0$} so the iterations in~\eqref{eq:gibbs_samping_algorithm} are nearly deterministic. In this case, a large $N_\text{iter}$ is required to obtain good APP estimates. 

To address stalling, one may use several samplers in parallel that are initialized with different bit strings. This makes stalling less likely but increases computational complexity. 
Alternatively, one can reduce confidence by increasing the variance of the channel likelihood; see~\cite[Sec. III]{senst2011rao} and~\cite{farhang2006markov,hansen_optimal_2009,auras_vlsi_2014}.
Consider sampling from a modified posterior PMF
\begin{align}
    \Omega(\gibbsvec) = \frac{1}{A_2} p(\mathbf{y} \lvert \gibbsvec, b(\bm{\hat{v}}^{s-1}),\hat{b}_1^{l-1}(\mathbf{v}_s))^{\frac{1}{\eta}}
    \label{eq:proposal_posterior}
\end{align}
where $\eta \in \left[1,\infty \right)$ is a confidence parameter  and $A_2$ is a normalization constant. For Gaussian channels, $\eta$ is a multiplicative factor for the noise variance.

The conditionals of~\eqref{eq:proposal_posterior} for Gibbs sampling are
\begin{equation}
\begin{aligned}
    & \Omega\big(\gibbsscalar_w \lvert \,\gibbsscalar^{w-1}(i), \gibbsscalar_{w+1}^W(i-1)\big) 
    = \frac{1}{A_3} \left(p\left(\mathbf{y} \lvert \,\mathbf{x}(\gibbsscalar_w)\,    \right) \right)^\frac{1}{\eta}
    \label{eq:gibbs_conditional_probability_eta}
\end{aligned}
\end{equation}
where $A_3$ is again a normalization constant. We may insert the modified conditionals~\eqref{eq:gibbs_conditional_probability_eta} into~\eqref{eq:gibbs_samping_algorithm}. As $N_\text{iter}\rightarrow\infty$  the distribution of  $\mathbf{\gibbsscalar}(N_\text{iter}), \gibbsvec(N_\text{iter+1}),\ldots$ tends to $\Omega(\gibbsvec)$.
We use importance sampling to estimate~\eqref{eq:app_pi} as 
\begin{align}
    \Phi(\gibbsscalar_w) \approx \frac{1}{N_\text{iter}} \sum_{i = 1}^{N_\text{iter}} \, \underbrace{\frac{\Phi(\gibbsvec_{[w]}(i))}{ \Omega(\gibbsvec_{[w]}(i))} }_\text{\small Importance weights}  \Phi(\gibbsscalar_w \lvert \gibbsvec_{[w]}(i))
    \label{eq:importance_sampling_estimation}
\end{align}
where the $\gibbsvec(i)$ are provided by the sampler and approximate $\Omega(\gibbsvec)$ and not the true posterior PMF. The importance weights resolve the mismatch. For $\eta=1$,~\eqref{eq:importance_sampling_estimation} becomes~\eqref{eq:app_est_simple_monte_c}.
For $\eta > 1$, one can avoid stalling at high SNR and the estimate~\eqref{eq:importance_sampling_estimation} is significantly better than~\eqref{eq:app_est_simple_monte_c} in general~\cite[Fig.~2]{senst2011rao}. Finally, one may optimize $\eta$ for every SNR by a line search with the corresponding rate expression as the cost function.

\subsection{Mismatched Decoding}
\label{sec:mismatched-decoding}
To further reduce complexity, we choose $\widetilde{K}$ to be smaller than the actual channel memory, i.e., we use mismatched decoding. Mismatched decoding is a standard tool for computing information rates for fiber-optic and wireless channels; see~\cite[Ex.~5.22]{Gallager68},~\cite{Divsalar78,Kaplan-Shamai-A93,Scarlett-FnT-20}, and also~\cite[Sec.~1.3-1.4]{Kramer-23} that reviews Gaussian auxiliary models.

The detector in~\cite{plabst2022achievable} replaces \eqref{eq:channel} with
\begin{align}
q(\bm{y}|\bm{u}) = \mathcal{N}\big(\bm{y} - \lvert \bm{\Psi'}
    \tilde{\bm{x}}' \rvert^{\circ 2}; \, \bm{\upmu}_\mathbf{Q},\, \covm{C}{\mathbf{QQ}}\big)
\label{eq:auxchannel}
\end{align}
where the matrix $\bm{\Psi'}$ is constructed as $\bm{\Psi}$, but with a smaller memory $\widetilde{K}$. We optimize $\bm{\upmu}_\mathbf{Q}$ and $\covm{\mathbf{C}}{\mathbf{QQ}}$ of the auxiliary channel~\eqref{eq:auxchannel} according to~\cite[Sec. III. C]{plabst2022achievable}. 

Let $q(\mathbf{y})=\sum_{\bm{u}} P(\bm{u})\,q(\bm{y}|\bm{u})$ and consider the rates
\begin{align}
    I_\text{$q$,$n$}(\mathbf{U};\mathbf{Y})
    & := \frac{1}{n} \E \left[ \log_2 \frac{q(\mathbf{Y}\lvert \mathbf{U})}{q(\mathbf{Y})}\right]
    \label{eq:mismatch_rate} \\
    I_\text{$q$,JDD} & := \lim_{n\to\infty} \frac{1}{n} I_\text{$q$,$n$}(\mathbf{U};\mathbf{Y})
    \label{eq:mismatch_rate_limit}
\end{align}
where the expectation is over the actual $P(\bm{u})p(\bm{y}|\bm{u})$. We then have (see~\cite[Sec. III. B]{plabst2022achievable} and~\cite{arnoldsimulationmi})
\begin{align}
    I_\text{$q$,$n$}(\mathbf{U};\mathbf{Y}) \le I_\text{$n$}(\mathbf{U};\mathbf{Y}),
    \quad I_\text{$q$,JDD} \le I_\text{JDD}.
\end{align}
We likewise compute the limiting rates $I_\text{$q$,SDD}$,
$I_\text{$q$,SIC}=I_\text{$q$,MSD}$, and $I_\text{$q$,b-SIC}$ as lower bounds on the respective $I_\text{SIC}=I_\text{MSD}$, $I_\text{SDD}$, and $I_\text{b-SIC}$ based on \eqref{eq:sic_inequality_alt}, \eqref{eq:sapp_inequality}, \eqref{eq:sic_mi_bits_finite_n} and \eqref{eq:rate_SIC_m-MSD_Gibbs}; see~\cite[Eq. (44)]{arnoldsimulationmi} or~\cite[Eq. (39)]{sadeghi2009optimization}. 

\section{Numerical Results}
\label{sec:numerical_results}
We simulate blocks with $n\ge \SI{20E3}{}$ transmit symbols over $L=\SI{0}{\kilo\meter}$ or $L=\SI{30}{\kilo\meter}$ of standard single-mode fiber without attenuation ($\SI{0}{dB/\kilo\meter}$). The symbol rate is $B=\SI{35}{\giga Baud}$ in the C~band at carrier wavelength $\SI{1550}{\nano\meter}$. The group velocity dispersion is $\beta_2=\SI{-2.168e-23}{\second^2\per\kilo\meter}$.
The average transmit power is
\begin{align}
    P_\text{tx} =  \frac{\E\big[ \lVert X(t) \rVert^2 \big]}{n\cdot T_\mathrm{s}}
\end{align} 
and the variance of the noise samples after filtering is \mbox{$N_0 B= 1$} so we define $\text{SNR}=P_\text{tx}$; see~\cite[Sec.~V]{plabst2022achievable}.
The spectral efficiency (SE) of FD-RC pulses is
\begin{align}
    \mathrm{SE} = R / (1 + \alpha) \quad \mathrm{[bit/s/Hz]}
\end{align}
where $R$ is the information rate in bpcu. 
We estimate $R$ using $I_\text{$q$,JDD}$, $I_\text{$q$,SIC}$, and $I_\text{$q$,b-SIC}$, that we compute via Monte Carlo simulation. The channel state $\bm{s}_0'$ (cf. Sec.~\ref{sec:mismatched-decoding}) is known at the receiver.

We use the $M$-ary modulation formats of~\cite{plabst2022achievable}:
\begin{itemize}
    \item unipolar $M$-PAM with $\mathcal{A}=\{0,1,\ldots,2^m-1\}$;
    \item bipolar $M$-ASK with $\mathcal{A}=\{\pm1,\pm3,\ldots \pm (2^m-1)\}$;
    \item complex-valued $M$-star-QAM ($M$-SQAM) with $\mathcal{A} = \left\{ \pm a, \pm \mathrm{j}a \mid a=1,2,\ldots,M/4  \right\}$.
\end{itemize}  
We approximate the combined filter $\psi(t)$ with a discrete filter $\psi_k$ with $K = N_\text{os} \cdot 101  +1$ taps, resulting in a system memory of $\widetilde{K} = 101$. This choice has $\psi_k$ containing $99.8\%$ of the energy of $\psi(t)$.

\subsection{SIC rates via FBA}
We use the FBA when the complexity is feasible.
\begin{figure*}[!t]
    \footnotesize
\begin{subfigure}[t]{0.33\textwidth}
\begin{tikzpicture}
\renewcommand{\baselinestretch}{1}
\pgfmathsetmacro\attshift{0}

\begin{axis}[%
MIGeneralStyle,ymax=2,
xmin=-2+\attshift,
xmax=10+\attshift,
ylabel={bpcu},
legend style={at={(0.99,0.01)},anchor=south east},
width = 5cm,
grid=both,
ytick={0,1,2},
]
\path[name path=axis] (axis cs:-5,0) -- (axis cs:15,0);
\addplot[PAM] table[x=power,y=rate,x expr=\thisrowno{0}+\attshift] {results/rates_bcjr/4-PAM/4-PAM_M=1_alpha=0.00_L=30km_mem=9_n=30000.txt}; 
\addplot[PAM,forget plot] table[x=power,y=rate,x expr=\thisrowno{0}+\attshift] {results/rates_bcjr/4-PAM/4-PAM_M=2_alpha=0.00_L=30km_mem=9_n=30000.txt}; 
\addplot[PAM,forget plot] table[x=power,y=rate,x expr=\thisrowno{0}+\attshift] {results/rates_bcjr/4-PAM/4-PAM_M=3_alpha=0.00_L=30km_mem=9_n=30000.txt}; 
\addplot[PAM,forget plot] table[x=power,y=rate,x expr=\thisrowno{0}+\attshift] {results/rates_bcjr/4-PAM/4-PAM_M=4_alpha=0.00_L=30km_mem=9_n=30000.txt}; 
\addplot[PAM,SP,name path=sp_pam] table[x=power,y=rate,x expr=\thisrowno{0}+\attshift] {results/sum-product/4-PAM/4-PAM_alpha=0.00_L=30km_mem=9.txt};

\addlegendentry{4-PAM, SIC}
\addlegendentry{4-PAM, JDD}

\draw[arrsingle]  (axis cs:2.95+\attshift,1.1) -- (axis cs:4.2+\attshift,0.85) node[xshift=0.3cm,below,font=black,opacitylabel,font=\footnotesize]{$S = 1,2,3,4$};

\node[rotate=45] at (axis cs: 5.4+\attshift,  1.3)[mycolor1,font=\footnotesize]{SDD}; 
\node[rotate=45] at (axis cs: 4.4+\attshift,1.5)[mycolor1,font=\footnotesize]{JDD}; 

\end{axis}
\end{tikzpicture}%
\vspace{0.2cm}
\end{subfigure}%
\begin{subfigure}[t]{0.33\textwidth}
\begin{tikzpicture}
\renewcommand{\baselinestretch}{1}
\pgfmathsetmacro\attshift{0}

\begin{axis}[%
MIGeneralStyle,ymax=2,
xmin=-2+\attshift,
xmax=10+\attshift,
ylabel={bpcu},
legend style={at={(0.99,0.01)},anchor=south east},
width = 5cm,,
grid=both,
ytick={0,1,2},
]

\addplot[ASK] table[x=power,y=rate,x expr=\thisrowno{0}+\attshift] {results/rates_bcjr/4-ASK/4-ASK_M=1_alpha=0.00_L=30km_mem=9_n=30000.txt}; 
\addplot[ASK,forget plot] table[x=power,y=rate,x expr=\thisrowno{0}+\attshift] {results/rates_bcjr/4-ASK/4-ASK_M=2_alpha=0.00_L=30km_mem=9_n=30000.txt}; 
\addplot[ASK,forget plot] table[x=power,y=rate,x expr=\thisrowno{0}+\attshift] {results/rates_bcjr/4-ASK/4-ASK_M=3_alpha=0.00_L=30km_mem=9_n=30000.txt}; 
\addplot[ASK,forget plot] table[x=power,y=rate,x expr=\thisrowno{0}+\attshift] {results/rates_bcjr/4-ASK/4-ASK_M=4_alpha=0.00_L=30km_mem=9_n=30000.txt}; 
\addplot[ASK,SP,name path=sp_ask] table[x=power,y=rate,x expr=\thisrowno{0}+\attshift] {results/sum-product/4-ASK/4-ASK_alpha=0.00_L=30km_mem=9.txt};

\addlegendentry{4-ASK, SIC}
\addlegendentry{4-ASK, JDD}

\draw[arrsingle]  (axis cs:2.25+\attshift,1.20) --(axis cs:4.3+\attshift,0.85) node[xshift=0.4cm,below,font=black,opacitylabel,font=\footnotesize]{$S = 1,2,3,4$};

\node[rotate=50] at (axis cs: 5.1+\attshift,  1.3)[mycolor6,font=\footnotesize]{SDD}; 
\node[rotate=45] at (axis cs: 3.2+\attshift,1.5)[mycolor6,font=\footnotesize]{JDD}; 

\end{axis}
\end{tikzpicture}%
\vspace{0.2cm}
\end{subfigure}
\begin{subfigure}[t]{0.33\textwidth}
\begin{tikzpicture}
\renewcommand{\baselinestretch}{1}
\pgfmathsetmacro\attshift{0}

\begin{axis}[%
MIGeneralStyle,ymax=2,
xmin=-2+\attshift,
xmax=10+\attshift,
ylabel={bpcu},
legend style={at={(0.99,0.01)},anchor=south east},
width = 5cm,,
grid=both,
ytick={0,1,2},
]

\addplot[QAM] table[x=power,y=rate,x expr=\thisrowno{0}+\attshift] {results/rates_bcjr/Q4/4-QPSK_M=1_alpha=0.00_L=30km_mem=9_n=30000_.txt}; 
\addplot[QAM,forget plot] table[x=power,y=rate,x expr=\thisrowno{0}+\attshift] {results/rates_bcjr/Q4/4-QPSK_M=2_alpha=0.00_L=30km_mem=9_n=30000_.txt}; 
\addplot[QAM,forget plot] table[x=power,y=rate,x expr=\thisrowno{0}+\attshift] {results/rates_bcjr/Q4/4-QPSK_M=3_alpha=0.00_L=30km_mem=9_n=30000_.txt}; 
\addplot[QAM,forget plot] table[x=power,y=rate,x expr=\thisrowno{0}+\attshift] {results/rates_bcjr/Q4/4-QPSK_M=4_alpha=0.00_L=30km_mem=9_n=30000_.txt}; 
\addplot[QAM,SP] table[x=power,y=rate,x expr=\thisrowno{0}+\attshift] {results/sum-product/4-QPSK_alpha=0.00_L=30km_mem=9.txt};

\addlegendentry{4-SQAM, SIC}
\addlegendentry{4-SQAM, JDD}

\draw[arrsingle]  (axis cs:2.4+\attshift,1.22) -- (axis cs:4.5+\attshift,0.85) node[xshift=0.4cm,below,font=black,opacitylabel,font=\footnotesize]{$S = 1,2,3,4$};

\node[rotate=50] at (axis cs: 5.2+\attshift,  1.3)[mycolor5,font=\footnotesize]{SDD}; 
\node[rotate=45] at (axis cs: 3.2+\attshift,1.5)[mycolor5,font=\footnotesize]{JDD}; 
\end{axis}
\end{tikzpicture}%
\vspace{0.2cm}
\end{subfigure}
\hfill
    \footnotesize
\begin{subfigure}[t]{0.33\textwidth}
\begin{tikzpicture}
\renewcommand{\baselinestretch}{1}
\pgfmathsetmacro\attshift{0}    

\begin{axis}[%
MIGeneralStyle,ymax=2.7+0.01,
xmin=-2+\attshift,
xmax=15+\attshift,
ylabel={bpcu},
legend style={at={(0.99,0.01)},anchor=south east},
width = 5cm
]
\path[name path=axis] (axis cs:-2,0) -- (axis cs:15,0);
\addplot[PAM] table[x=power,y=rate, x expr=\thisrowno{0}+\attshift] {results/rates_bcjr/Q8/8-PAM_M=1_alpha=0.00_L=30km_mem=7.txt}; 
\addplot[PAM,forget plot] table[x=power,y=rate, x expr=\thisrowno{0}+\attshift] {results/rates_bcjr/Q8/8-PAM_M=2_alpha=0.00_L=30km_mem=7.txt}; 
\addplot[PAM,forget plot] table[x=power,y=rate, x expr=\thisrowno{0}+\attshift] {results/rates_bcjr/Q8/8-PAM_M=3_alpha=0.00_L=30km_mem=7.txt}; 
\addplot[PAM,forget plot] table[x=power,y=rate, x expr=\thisrowno{0}+\attshift] {results/rates_bcjr/Q8/8-PAM_M=4_alpha=0.00_L=30km_mem=7.txt}; 
\addplot[PAM,SP] table[x=power,y=rate, x expr=\thisrowno{0}+\attshift] {results/sum-product/8-PAM/8-PAM_alpha=0.00_L=30km_mem=7.txt};

\addlegendentry{8-PAM, SIC}
\addlegendentry{8-PAM, JDD}

\draw[arrsingle]  (axis cs:4.7+\attshift,1.5) -- (axis cs:7.2+\attshift,1.15) node[xshift=0.3cm,below,font=black,opacitylabel,font=\footnotesize]{$S = 1,2,3,4$};

\node[rotate=45] at (axis cs: 7.9+\attshift,  1.5)[mycolor1,font=\footnotesize]{SDD}; 
\node[rotate=50] at (axis cs: 6+\attshift,1.8)[mycolor1,font=\footnotesize]{JDD}; 

\end{axis}
\end{tikzpicture}%
 \end{subfigure}
  \begin{subfigure}[t]{0.33\textwidth}
\begin{tikzpicture}
\renewcommand{\baselinestretch}{1}
\pgfmathsetmacro\attshift{0}

\begin{axis}[%
MIGeneralStyle,ymax=2.7+0.01,
xmin=-2+\attshift,
xmax=15+\attshift,
ylabel={bpcu},
legend style={at={(0.99,0.01)},anchor=south east},
width = 5cm
]

\addplot[ASK] table[x=power,y=rate, x expr=\thisrowno{0}+\attshift] {results/rates_bcjr/Q8/8-ASK_M=1_alpha=0.00_L=30km_mem=7.txt}; 
\addplot[ASK,forget plot] table[x=power,y=rate, x expr=\thisrowno{0}+\attshift] {results/rates_bcjr/Q8/8-ASK_M=2_alpha=0.00_L=30km_mem=7.txt}; 
\addplot[ASK,forget plot] table[x=power,y=rate, x expr=\thisrowno{0}+\attshift] {results/rates_bcjr/Q8/8-ASK_M=3_alpha=0.00_L=30km_mem=7.txt}; 
\addplot[ASK,forget plot] table[x=power,y=rate, x expr=\thisrowno{0}+\attshift] {results/rates_bcjr/Q8/8-ASK_M=4_alpha=0.00_L=30km_mem=7.txt}; 
\addplot[ASK,SP] table[x=power,y=rate, x expr=\thisrowno{0}+\attshift] {results/sum-product/8-ASK/8-ASK_alpha=0.00_L=30km_mem=7.txt};

\draw[arrsingle]  (axis cs:4.3+\attshift,1.6) -- (axis cs:7.2+\attshift,1.15) node[xshift=0.4cm,below,font=black,opacitylabel,font=\footnotesize]{$S = 1,2,3,4$};

\node[rotate=45] at (axis cs: 8.5+\attshift,  1.5)[mycolor6,font=\footnotesize]{SDD}; 
\node[rotate=50] at (axis cs: 4.5+\attshift,1.8)[mycolor6,font=\footnotesize]{JDD}; 

\addlegendentry{8-ASK, SIC}
\addlegendentry{8-ASK, JDD}

\end{axis}
\end{tikzpicture}%
 \end{subfigure}%
  \begin{subfigure}[t]{0.33\textwidth}
\begin{tikzpicture}
\renewcommand{\baselinestretch}{1}
\pgfmathsetmacro\attshift{0}

\begin{axis}[%
MIGeneralStyle,ymax=2.7+0.01,
xmin=-2+\attshift,
xmax=15+\attshift,
ylabel={bpcu},
legend style={at={(0.99,0.01)},anchor=south east},
width = 5cm
]

\addplot[QAM] table[x=power,y=rate, x expr=\thisrowno{0}+\attshift] {results/rates_bcjr/Q8/8-SQAM_M=1_alpha=0.00_L=30km_mem=7.txt}; 
\addplot[QAM,forget plot] table[x=power,y=rate, x expr=\thisrowno{0}+\attshift] {results/rates_bcjr/Q8/8-SQAM_M=2_alpha=0.00_L=30km_mem=7.txt}; 
\addplot[QAM,forget plot] table[x=power,y=rate, x expr=\thisrowno{0}+\attshift] {results/rates_bcjr/Q8/8-SQAM_M=3_alpha=0.00_L=30km_mem=7.txt}; 
\addplot[QAM,forget plot] table[x=power,y=rate, x expr=\thisrowno{0}+\attshift] {results/rates_bcjr/Q8/8-SQAM_M=4_alpha=0.00_L=30km_mem=7.txt}; 
\addplot[QAM,SP] table[x=power,y=rate, x expr=\thisrowno{0}+\attshift] {results/sum-product/8-QAM_alpha=0.00_L=30km_mem=7.txt};

\draw[arrsingle]  (axis cs:4.3+\attshift,1.6) -- (axis cs:7.2+\attshift,1.15) node[xshift=0.8cm,below,font=black,opacitylabel,font=\footnotesize]{$S = 1,2,3,4$};

\node[rotate=45] at (axis cs: 8.5+\attshift,  1.5)[mycolor5,font=\footnotesize]{SDD}; 
\node[rotate=50] at (axis cs: 4.3+\attshift,1.8)[mycolor5,font=\footnotesize]{JDD};

\addlegendentry{8-SQAM, SIC}
\addlegendentry{8-SQAM, JDD}

\end{axis}
\end{tikzpicture}%
 \end{subfigure}
 \hfill
    \caption{Rates $I_{q,\text{SIC}}$ for $S=1,2,3,4$ stages and $I_{q,\text{JDD}}$, $L = \SI{30}{\kilo\meter}$, $\widetilde{K}=9$ and $\widetilde{K}=7$ for $M=4$ and $M=8$, respectively.}
    \label{fig:sic_rates_4_8_ary}
\end{figure*}
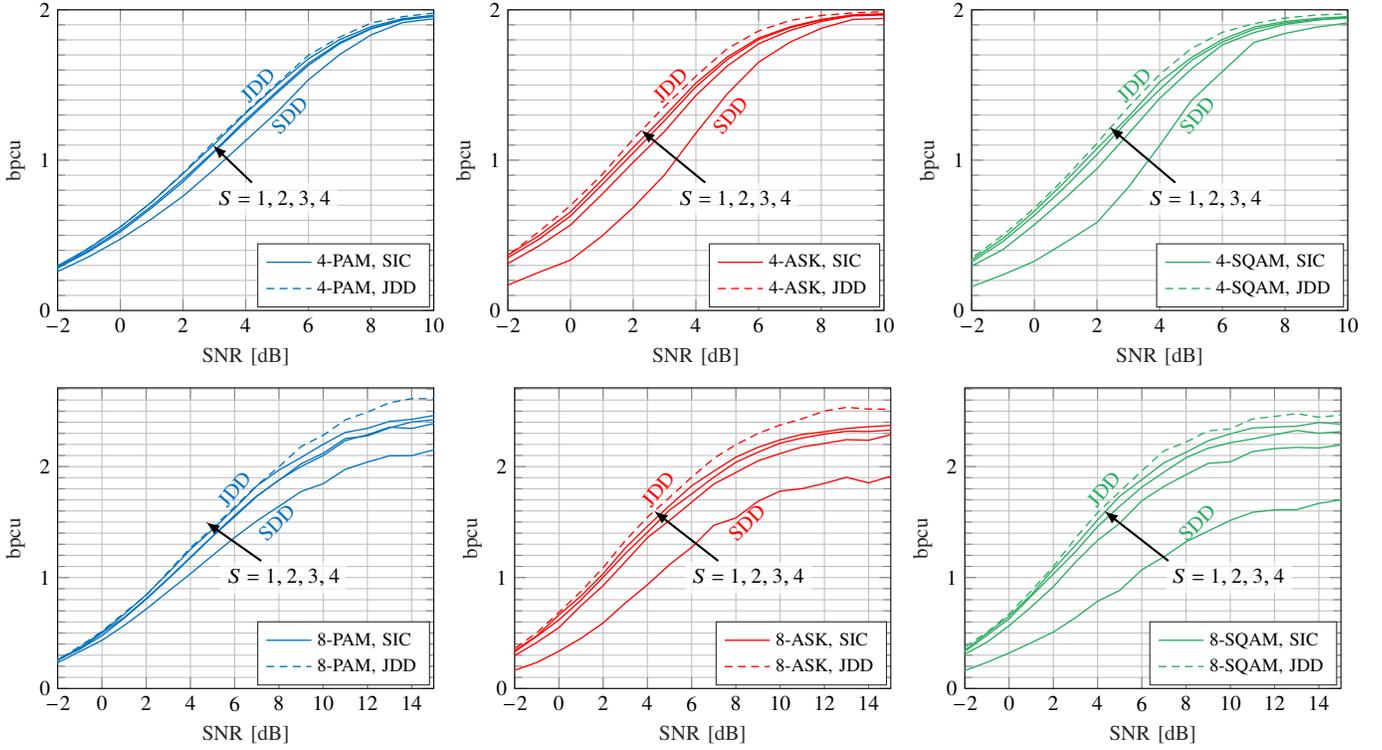
Fig.~\ref{fig:sic_rates_4_8_ary} shows the rates $I_\text{$q$,JDD}$ and $I_\text{$q$,SIC}$ for $L = \SI{30}{\kilo\meter}$ of fiber and with sinc pulses. The rates increase with the number of SIC stages $S$. Observe that $\widetilde{K} = 9$ suffices to achieve nearly $\SI{2}{bpcu}$ for all $4$-ary alphabets and large SNR; see also~\cite[Fig. 8a]{plabst2022achievable}. For $M=4$, SDD achieves good rates for large SNR; see also~\cite[Fig. 1]{plabst2022achievable} for BPSK. For medium SNR, SDD exhibits larger losses for ASK and SQAM. For $M=8$ and $\widetilde{K} = 7$, the rates are limited by interference and saturate before \SI{3}{bpcu}. Increasing $\widetilde{K}$ would thus achieve higher rates. Likewise, SDD shows large losses for ASK and 8-SQAM for medium SNR due to the small auxiliary memory. 

Fig.~\ref{fig:sic_rates_4_8_ary} shows that SIC approaches the JDD rates for large $S$.
The performance with $S=4$ agrees with JDD at low-to-medium SNR.  For large $S$, the energy gains for ASK, PAM, and SQAM are similar to those in~\cite{plabst2022achievable}. 

\subsection{Polar-Coded SDD and SIC with the FBA} \label{sec:polar}
Practical communication systems can achieve the SIC rates in Fig.~\ref{fig:sic_rates_4_8_ary}. To show this, we use polar codes~\cite{arikan_channel_2009} combined with MLC/MSD~\cite{seidl_polar_2013}. 
We further use successive cancellation list (SCL) decoding~\cite{tal_list_2015} that passes a list of the most likely codewords between stages. 
The detector of the next stage uses this list to generate a list of APPs for the next decoder~\cite{prinz_successive_2018,karakchieva_joint_2019}, i.e., we pass a list of possible bit strings to the next stage instead of only one hard-decision string.

We simulate $10^{6}$ frames that each have $1024$ \mbox{4-PAM/ASK} symbols. For MSD with 4-ASK/PAM (2 bit-levels), we use $2S$ binary polar codes of length $2048/(2S)$ bits each. We set the rate~\eqref{eq:sic_information_rate} to $R=\SI{1}{bpcu}$, i.e., there are $k=1024$ data bits per transmitted frame. The SCL list size is eight, and we use an outer 16-bit cyclic redundancy check (CRC) code; see~\cite{tal_list_2015}.
The polar codes are designed using Monte Carlo methods~\cite{arikan_channel_2009,bocherer2017efficient}, and the code rate of all component codes ($S$ SIC levels, $m$ bit levels) is simultaneously optimized. One can find the least reliable bit positions, i.e., the frozen bits, by decoding many frames to estimate the reliability of every bit position.

\begin{figure*}[!t] %
    \begin{subfigure}[t]{0.5\textwidth}
    \centering
        \begin{tikzpicture}
\footnotesize
\renewcommand{\baselinestretch}{1}
\pgfmathsetmacro\rcosrolloffa{0.2}    
\pgfmathsetmacro\attshift{0}

\begin{axis}[%
MIGeneralStyle,
xmax=4+\attshift,
xmin=1.8+\attshift),
ymin=0.9/(1+\rcosrolloffa),
ymax=1.2/(1+\rcosrolloffa),
ylabel={SE [bit/s/Hz]},
xtick distance={0.25},
width=7.5cm,
height=\gridfigureheight,
]

\addplot[ASK,name path global=ask1] table[x=power,y=se, x expr=\thisrowno{0}+\attshift] {results/rates_bcjr/4-ASK/se_4-ASK_M=1_alpha=0.20_L=30km_mem=5_n=100000.txt}; 
\addplot[ASK,forget plot,name path global=ask2] table[x=power,y=se, x expr=\thisrowno{0}+\attshift] {results/rates_bcjr/4-ASK/se_4-ASK_M=2_alpha=0.20_L=30km_mem=5_n=100000.txt}; 

\addplot[ASK,forget plot,name path global=ask4] table[x=power,y=se, x expr=\thisrowno{0}+\attshift] {results/rates_bcjr/4-ASK/se_4-ASK_M=4_alpha=0.20_L=30km_mem=5_n=100000.txt}; 

\addplot[PAM,name path global=pam1,densely dashed] table[x=power,y=se, x expr=\thisrowno{0}+\attshift] {results/rates_bcjr/4-PAM/se_4-PAM_M=1_alpha=0.20_L=30km_mem=5_n=100000.txt}; 
\addplot[PAM,forget plot,name path global=pam2,densely dashed] table[x=power,y=se, x expr=\thisrowno{0}+\attshift] {results/rates_bcjr/4-PAM/se_4-PAM_M=2_alpha=0.20_L=30km_mem=5_n=100000.txt}; 

\addplot[PAM,forget plot,name path global=pam4,densely dashed] table[x=power,y=se, x expr=\thisrowno{0}+\attshift] {results/rates_bcjr/4-PAM/se_4-PAM_M=4_alpha=0.20_L=30km_mem=5_n=100000.txt};

\addlegendentry{4-ASK, $S=1,2,4$}
\addlegendentry{4-PAM, $S=1,2,4$}

\path[name path global=line] (axis cs:\pgfkeysvalueof{/pgfplots/xmin},0.84) -- (axis cs: \pgfkeysvalueof{/pgfplots/xmax},0.84);
\path[name intersections={of=line and pam4, name=p1}, name intersections={of=line and pam1, name=p2}];
\draw[arr,thick] let \p1=(p1-1), \p2=(p2-1) in (p1-1) -- ([xshift=-0.05cm]p2-1) node [above,midway,xshift=0.45cm,opacitylabel] {%
	\pgfplotsconvertunittocoordinate{x}{\x1}%
	\pgfplotscoordmath{x}{datascaletrafo inverse to fixed}{\pgfmathresult}%
	\edef\valueA{\pgfmathresult}%
	\pgfplotsconvertunittocoordinate{x}{\x2}%
	\pgfplotscoordmath{x}{datascaletrafo inverse to fixed}{\pgfmathresult}%
	\pgfmathparse{\pgfmathresult - \valueA}%
	\pgfmathprintnumber{\pgfmathresult} dB
};

\path[name path global=line] (axis cs:\pgfkeysvalueof{/pgfplots/xmin},0.82) -- (axis cs: \pgfkeysvalueof{/pgfplots/xmax},0.82);
\path[name intersections={of=line and ask4, name=p1}, name intersections={of=line and pam1, name=p2}];
\draw[arr,thick] let \p1=(p1-1), \p2=(p2-1) in (p1-1) -- ([xshift=-0.05cm]p2-1) node [below,midway,xshift=0.5cm,opacitylabel] {%
	\pgfplotsconvertunittocoordinate{x}{\x1}%
	\pgfplotscoordmath{x}{datascaletrafo inverse to fixed}{\pgfmathresult}%
	\edef\valueA{\pgfmathresult}%
	\pgfplotsconvertunittocoordinate{x}{\x2}%
	\pgfplotscoordmath{x}{datascaletrafo inverse to fixed}{\pgfmathresult}%
	\pgfmathparse{\pgfmathresult - \valueA}%
	\pgfmathprintnumber{\pgfmathresult} dB
};

\path[name path global=line] (axis cs:\pgfkeysvalueof{/pgfplots/xmin},0.84) -- (axis cs: \pgfkeysvalueof{/pgfplots/xmax},0.84);
\path[name intersections={of=line and ask4, name=p1}, name intersections={of=line and pam4, name=p2}];
\draw[arr,thick] let \p1=(p1-1), \p2=(p2-1) in (p1-1) -- ([xshift=-0.05cm]p2-1) node [above,midway,xshift=-0.1cm,opacitylabel] {%
	\pgfplotsconvertunittocoordinate{x}{\x1}%
	\pgfplotscoordmath{x}{datascaletrafo inverse to fixed}{\pgfmathresult}%
	\edef\valueA{\pgfmathresult}%
	\pgfplotsconvertunittocoordinate{x}{\x2}%
	\pgfplotscoordmath{x}{datascaletrafo inverse to fixed}{\pgfmathresult}%
	\pgfmathparse{\pgfmathresult - \valueA}%
	\pgfmathprintnumber{\pgfmathresult} dB
};

\path[name path global=line] (axis cs:\pgfkeysvalueof{/pgfplots/xmin},0.86) -- (axis cs: \pgfkeysvalueof{/pgfplots/xmax},0.86);
\path[name intersections={of=line and ask2, name=p1}, name intersections={of=line and pam2, name=p2}];
\draw[arr,thick] let \p1=(p1-1), \p2=(p2-1) in (p1-1) -- ([xshift=-0.05cm]p2-1) node [above,midway,opacitylabel] {%
	\pgfplotsconvertunittocoordinate{x}{\x1}%
	\pgfplotscoordmath{x}{datascaletrafo inverse to fixed}{\pgfmathresult}%
	\edef\valueA{\pgfmathresult}%
	\pgfplotsconvertunittocoordinate{x}{\x2}%
	\pgfplotscoordmath{x}{datascaletrafo inverse to fixed}{\pgfmathresult}%
	\pgfmathparse{\pgfmathresult - \valueA}%
	\pgfmathprintnumber{\pgfmathresult} dB
};

\end{axis}
\end{tikzpicture}%
        \caption{Achievable rates around $\mathrm{SE} \approx 0.83 \,\text{bit/s/Hz}$.}
        \label{fig:polar_rates}
    \end{subfigure}%
    \hfill%
    \begin{subfigure}[t]{0.5\textwidth}
    \centering
        \begin{tikzpicture}
\footnotesize
\renewcommand{\baselinestretch}{1}
\pgfmathsetmacro\attshift{0}

\begin{axis}[%
FERGeneralStyle,
xmin=3+\attshift,
xmax=7+\attshift,
ymode=log,
ymax=1,
ymin=3e-4,
width=7.5cm,
height=\gridfigureheight,
]

\addplot[ASK,name path global=ask1,mark=diamond] table[x=power,y=fer, x expr=\thisrowno{0}+\attshift] {results/coded-results/pcm-sic_4-ASK_M=1_n=1024_R=1.00_listsize=8_16-CRC_P-design=5.00dB_alpha=0.20_L=30km_mem=5_is-systematic=1.txt}; 
\addplot[ASK,forget plot,name path global=ask2,mark=diamond] table[x=power,y=fer, x expr=\thisrowno{0}+\attshift] {results/coded-results/pcm-sic_4-ASK_M=2_n=1024_R=1.00_listsize=8_16-CRC_P-design=4.00dB_alpha=0.20_L=30km_mem=5_is-systematic=1.txt}; 

\addplot[ASK,forget plot,name path global=ask4,mark=diamond] table[x=power,y=fer, x expr=\thisrowno{0}+\attshift] {results/coded-results/pcm-sic_4-ASK_M=4_n=1024_R=1.00_listsize=8_16-CRC_P-design=4.00dB_alpha=0.20_L=30km_mem=5_is-systematic=1.txt}; 

\addplot[PAM,name path global=pam1, densely dashed,mark=asterisk,mark options={solid}] table[x=power,y=fer, x expr=\thisrowno{0}+\attshift] {results/coded-results/pcm-sic_4-PAM_M=1_n=1024_R=1.00_listsize=8_16-CRC_P-design=4.00dB_alpha=0.20_L=30km_mem=5_is-systematic=1.txt}; 
\addplot[PAM,forget plot,name path global=pam2,densely dashed,mark=asterisk,mark options={solid}] table[x=power,y=fer, x expr=\thisrowno{0}+\attshift] {results/coded-results/pcm-sic_4-PAM_M=2_n=1024_R=1.00_listsize=8_16-CRC_P-design=4.00dB_alpha=0.20_L=30km_mem=5_is-systematic=1.txt}; 
\addplot[PAM,forget plot,name path global=pam4,densely dashed,mark=asterisk,mark options={solid}] table[x=power,y=fer, x expr=\thisrowno{0}+\attshift] {results/coded-results/pcm-sic_4-PAM_M=4_n=1024_R=1.00_listsize=8_16-CRC_P-design=4.00dB_alpha=0.20_L=30km_mem=5_is-systematic=1.txt};

\addlegendentry{4-ASK, $S=1,2,4$}
\addlegendentry{4-PAM, $S=1,2,4$}

\path[name path global=line] (axis cs:\pgfkeysvalueof{/pgfplots/xmin},1.2e-3) -- (axis cs: \pgfkeysvalueof{/pgfplots/xmax},1.2e-3);
\path[name intersections={of=line and pam4, name=p1}, name intersections={of=line and pam1, name=p2}];
\draw[arr,thick] let \p1=(p1-1), \p2=(p2-1) in (p1-1) -- ([xshift=-0.05cm]p2-1) node [above,midway,xshift=0.3cm,opacitylabel] {%
	\pgfplotsconvertunittocoordinate{x}{\x1}%
	\pgfplotscoordmath{x}{datascaletrafo inverse to fixed}{\pgfmathresult}%
	\edef\valueA{\pgfmathresult}%
	\pgfplotsconvertunittocoordinate{x}{\x2}%
	\pgfplotscoordmath{x}{datascaletrafo inverse to fixed}{\pgfmathresult}%
	\pgfmathparse{\pgfmathresult - \valueA}%
	\pgfmathprintnumber{\pgfmathresult} dB
};

\path[name path global=line] (axis cs:\pgfkeysvalueof{/pgfplots/xmin},9e-4) -- (axis cs: \pgfkeysvalueof{/pgfplots/xmax},9e-4);
\path[name intersections={of=line and ask4, name=p1}, name intersections={of=line and pam1, name=p2}];
\draw[arr,thick] let \p1=(p1-1), \p2=(p2-1) in (p1-1) -- ([xshift=-0.05cm]p2-1) node [below,midway,fill=white,opacitylabel] {%
	\pgfplotsconvertunittocoordinate{x}{\x1}%
	\pgfplotscoordmath{x}{datascaletrafo inverse to fixed}{\pgfmathresult}%
	\edef\valueA{\pgfmathresult}%
	\pgfplotsconvertunittocoordinate{x}{\x2}%
	\pgfplotscoordmath{x}{datascaletrafo inverse to fixed}{\pgfmathresult}%
	\pgfmathparse{\pgfmathresult - \valueA}%
	\pgfmathprintnumber{\pgfmathresult} dB
};

\path[name path global=line] (axis cs:\pgfkeysvalueof{/pgfplots/xmin},1.2e-3) -- (axis cs: \pgfkeysvalueof{/pgfplots/xmax},1.2e-3);
\path[name intersections={of=line and ask4, name=p1}, name intersections={of=line and pam4, name=p2}];
\draw[arr,thick] let \p1=(p1-1), \p2=(p2-1) in (p1-1) -- ([xshift=-0.05cm]p2-1) node [above,midway,xshift=-0.9cm,yshift=0.0cm,opacitylabel] {%
	\pgfplotsconvertunittocoordinate{x}{\x1}%
	\pgfplotscoordmath{x}{datascaletrafo inverse to fixed}{\pgfmathresult}%
	\edef\valueA{\pgfmathresult}%
	\pgfplotsconvertunittocoordinate{x}{\x2}%
	\pgfplotscoordmath{x}{datascaletrafo inverse to fixed}{\pgfmathresult}%
	\pgfmathparse{\pgfmathresult - \valueA}%
	\pgfmathprintnumber{\pgfmathresult} dB
};

\path[name path global=line] (axis cs:\pgfkeysvalueof{/pgfplots/xmin},1.6e-3) -- (axis cs: \pgfkeysvalueof{/pgfplots/xmax},1.6e-3);
\path[name intersections={of=line and ask2, name=p1}, name intersections={of=line and pam2, name=p2}];
\draw[arr,thick] let \p1=(p1-1), \p2=(p2-1) in (p1-1) -- ([xshift=-0.05cm]p2-1) node [above,midway,yshift=0.1cm,opacitylabel] {%
	\pgfplotsconvertunittocoordinate{x}{\x1}%
	\pgfplotscoordmath{x}{datascaletrafo inverse to fixed}{\pgfmathresult}%
	\edef\valueA{\pgfmathresult}%
	\pgfplotsconvertunittocoordinate{x}{\x2}%
	\pgfplotscoordmath{x}{datascaletrafo inverse to fixed}{\pgfmathresult}%
	\pgfmathparse{\pgfmathresult - \valueA}%
	\pgfmathprintnumber{\pgfmathresult} dB
};

\end{axis}
\end{tikzpicture}%
        \caption{FER at a $\text{SE} \approx 0.83\,\text{bit/s/Hz}$ for polar coded modulation.} \label{fig:polar_fer}
    \end{subfigure}  
    \caption{SE based on $I_\text{$q$,SIC}$ and polar-coded FER for 4-ASK/PAM, $\widetilde{K}=5$, FD-RC $(\alpha=0.2)$ and $L = \SI{30}{\kilo\meter}$.}
    \label{fig:polar_results}
\end{figure*}
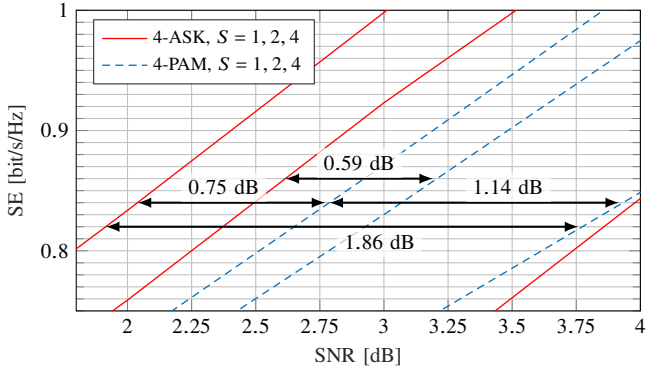
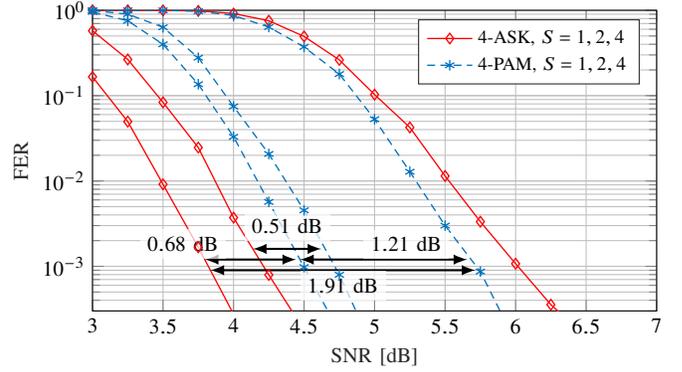

Fig.~\ref{fig:polar_rates} shows $I_\text{$q$,SIC}$ for
$\widetilde{K}=5$ and a FD-RC pulse with $\alpha=0.2$ so ${\rm SE} \approx \SI{0.83}{bit/s/Hz}$. For SDD, the  4-ASK and 4-PAM rates almost coincide. Using SIC with $S=2,4$, ASK gains approximately $\SI{0.59}{dB}$ and $\SI{0.75}{dB}$ in SNR over PAM, respectively, at \SI{0.83}{bit/s/Hz}. Comparing SIC with $S=4$ to SDD ($S=1$), PAM and ASK gain \SI{1.14}{dB} and \SI{1.86}{dB} in SNR, respectively. 
Fig.~\ref{fig:polar_fer} shows the FERs for which the SNR gains closely match those of Fig.~\ref{fig:polar_rates} at FER $10^{-3}$. 

We remark that fiber-optic applications often require end-to-end FERs below $10^{-15}$. One can achieve this by serially concatenating a soft-decision inner code, such as a low-density parity-check or trellis code, with an interleaver and a powerful outer code to reduce the FER to the desired level, e.g., a Reed-Solomon code, a Bose–Chaudhuri–Hocquenghem (BCH) code, or a staircase code; see~\cite{ITU-04,Mizuochi-09,Magarini-10,Smith-12}. Serial concatenation with a soft-decision inner code is also part of the 5G wireless standard, e.g., with a polar inner code and a CRC outer code as described above; see~\cite{Kumar-23}.

\subsection{SIC rates via Gibbs sampling}
This section presents results for $I_\text{$q$,b-SIC}$ where the APPs are approximated by Gibbs sampling. The results are compared to $I_\text{$q$,SIC}$ where the APPs are computed via the FBA. To avoid stalling at high SNR, we use $N_\text{par} = 20$ parallel samplers and perform $N_\mathrm{iter}=60$ iterations per sampler. The first few (here, ten) iterations are discarded (the so-called burn-in period). For every SNR, we use a line search with a training set of $\SI{10e3}{}$ symbols to optimize the confidence parameter $\eta$ with respect to $I_\text{$q$,b-SIC}$. We use a binary reflected Gray code to label the elements in the alphabet~$\mathcal{A}$.

Fig.~\ref{fig:bcjr_gibbs_comparison} shows results for $L=\SI{30}{\kilo\meter}$ and sinc pulses and plots the SIC rates $I_\text{$q$,b-SIC}$ for 4-ASK, 8-ASK, and $S=1,\ldots,4$, where the APPs are estimated via sampling.
\begin{figure*}[!t] %
    \begin{subfigure}[t]{0.49\textwidth}
            \centering
            \begin{tikzpicture}
\renewcommand{\baselinestretch}{1}

\begin{axis}[%
MIGeneralStyle,ymax=2,
xmin=-2,
xmax=10,
ylabel={bpcu},
legend style={at={(0.99,0.01)},anchor=south east},
width=7.5cm,
ytick={0,1,2},
]

\addplot[ASK,name path=sp_ask,dashed,forget plot] table[x=power,y=rate] {results/sum-product/4-ASK/4-ASK_alpha=0.00_L=30km_mem=9.txt}; 
\addplot[ASK] table[x=power,y=rate] {results/rates_bcjr/4-ASK/4-ASK_M=1_alpha=0.00_L=30km_mem=9_n=30000.txt}; 
\addplot[ASK,forget plot] table[x=power,y=rate] {results/rates_bcjr/4-ASK/4-ASK_M=2_alpha=0.00_L=30km_mem=9_n=30000.txt}; 
\addplot[ASK,forget plot] table[x=power,y=rate] {results/rates_bcjr/4-ASK/4-ASK_M=3_alpha=0.00_L=30km_mem=9_n=30000.txt}; 
\addplot[ASK,forget plot] table[x=power,y=rate] {results/rates_bcjr/4-ASK/4-ASK_M=4_alpha=0.00_L=30km_mem=9_n=30000.txt};

\addplot[ASKGibbs] table[x=power,y=rate] {results/rates_gibbs/Q4/senst__Gibbs__4-ASK__S=1__alpha=0.00__L=30km__mem=9__it=50__par=20__n=20000__burn=10.txt};
\addplot[ASKGibbs,forget plot] table[x=power,y=rate] {results/rates_gibbs/Q4/senst__Gibbs__4-ASK__S=2__alpha=0.00__L=30km__mem=9__it=50__par=20__n=20000__burn=10.txt};
\addplot[ASKGibbs,forget plot] table[x=power,y=rate] {results/rates_gibbs/Q4/senst__Gibbs__4-ASK__S=3__alpha=0.00__L=30km__mem=9__it=50__par=20__n=20000__burn=10.txt};
\addplot[ASKGibbs,forget plot] table[x=power,y=rate] {results/rates_gibbs/Q4/senst__Gibbs__4-ASK__S=4__alpha=0.00__L=30km__mem=9__it=50__par=20__n=20000__burn=10.txt};

\draw[arrsingle,shadowed]  (axis cs:2.7,1.26) -- (axis cs:4.5,0.85) node[below,font=black,font=\footnotesize,opacitylabel]{$S = 1,\ldots,4$};

\node[rotate=40] at (axis cs: 5.0,  1.3)[mycolor6,font=\footnotesize]{SDD}; 
\node[rotate=35] at (axis cs:4.2,1.7)[mycolor6,font=\footnotesize]{JDD (FBA)}; 

\addlegendentry{FBA, 4-ASK, $\tilde{N} = 9$}
\addlegendentry{Gibbs, 4-ASK, $\tilde{N} = 9$}

\end{axis}
\end{tikzpicture}%
    \end{subfigure}%
    \hfill%
    \begin{subfigure}[t]{0.49\textwidth}
            \centering 
            \begin{tikzpicture}
\renewcommand{\baselinestretch}{1}

\begin{axis}[%
MIGeneralStyle,ymax=3+0.01,
xmin=-2,
xmax=15,
ylabel={bpcu},
legend style={at={(0.99,0.01)},anchor=south east},
width=7.5cm,
]

\addplot[ASK,dashed,forget plot] table[x=power,y=rate] {results/sum-product/8-ASK/8-ASK_alpha=0.00_L=30km_mem=7.txt};

\addplot[ASK] table[x=power,y=rate] {results/rates_bcjr/Q8/8-ASK_M=1_alpha=0.00_L=30km_mem=7.txt}; 
\addplot[ASK,forget plot] table[x=power,y=rate] {results/rates_bcjr/Q8/8-ASK_M=2_alpha=0.00_L=30km_mem=7.txt}; 
\addplot[ASK,forget plot] table[x=power,y=rate] {results/rates_bcjr/Q8/8-ASK_M=3_alpha=0.00_L=30km_mem=7.txt}; 
\addplot[ASK,forget plot] table[x=power,y=rate] {results/rates_bcjr/Q8/8-ASK_M=4_alpha=0.00_L=30km_mem=7.txt}; 

\addplot[ASKGibbs] table[x=power,y=rate] {results/rates_gibbs/Q8/senst__Gibbs__8-ASK__S=1__alpha=0.00__L=30km__mem=7__it=50__par=20__n=20000__burn=10.txt};
\addplot[ASKGibbs,forget plot] table[x=power,y=rate] {results/rates_gibbs/Q8/senst__Gibbs__8-ASK__S=2__alpha=0.00__L=30km__mem=7__it=50__par=20__n=20000__burn=10.txt};
\addplot[ASKGibbs,forget plot] table[x=power,y=rate] {results/rates_gibbs/Q8/senst__Gibbs__8-ASK__S=3__alpha=0.00__L=30km__mem=7__it=50__par=20__n=20000__burn=10.txt};
\addplot[ASKGibbs,forget plot] table[x=power,y=rate] {results/rates_gibbs/Q8/senst__Gibbs__8-ASK__S=4__alpha=0.00__L=30km__mem=7__it=50__par=20__n=20000__burn=10.txt};

\addlegendentry{FBA, 8-ASK, $\tilde{N} = 7$}
\addlegendentry{Gibbs, 8-ASK, $\tilde{N} = 7$}

\draw[arrsingle,shadowed]  (axis cs:5.3,1.75) -- (axis cs:7.1,1.25) node[xshift=0.4cm,below,font=black,opacitylabel,font=\footnotesize]{$S = 1,\ldots,4$};

\node[rotate=30] at (axis cs: 8.5,  1.5)[mycolor6,font=\footnotesize]{SDD}; 
\node[rotate=28] at (axis cs: 7.3,2.3)[mycolor6,font=\footnotesize]{JDD (FBA)}; 

\end{axis}
\end{tikzpicture}%
    \end{subfigure}  
        \caption{Comparison of $I_\text{$q$,b-SIC}$ (Gibbs sampling) and $I_\text{$q$,SIC}$ (FBA) for sinc pulses and $L = \SI{30}{\kilo\meter}$.}
        \label{fig:bcjr_gibbs_comparison}
\end{figure*}
Using~\eqref{eq:bsic_inequality}, we have $I_\text{SIC} \le I_\text{b-SIC}$, and for a sufficiently large $\widetilde{K}$ this should be reflected in Fig.~\ref{fig:bcjr_gibbs_comparison}. The figure shows that $I_\text{$q$,SIC} \le I_\text{$q$,b-SIC}$ for $S=1$ and low SNR. The sampler still stalls at high SNR, and the rates saturate. PAM and SQAM behave similarly when comparing $I_\text{$q$,SIC}$ computed via the FBA to $I_\text{$q$,b-SIC}$ computed via Gibbs sampling. These results are omitted.
As $S$ increases, the rates $I_\text{$q$,b-SIC}$ via Gibbs sampling approach the JDD curve at low and medium SNR, but not at high SNR. Remedies for this behavior include more iterations and parallel samplers.

\begin{figure}[!t] %
    \centering
    \begin{tikzpicture}
\renewcommand{\baselinestretch}{1}
\pgfmathsetmacro\rcosrolloffa{0.2}

\begin{axis}[%
MIGeneralStyle,
xmax=19,
xmin=0,
ylabel={$I_{q,\text{b-SIC}}$ [bpcu]},
ymax=5,
ytick distance=1,
legend columns=3,
legend style = {at={(1, 0)}, anchor=south east, style={column sep=0.05cm}}, 
minor y tick num=4,
width=7.5cm,
legend style={at={(0.5,1.04)},anchor=south}
]

\addplot[PAM, line width=0.7pt,line width=0.9pt,name path global=pam8,mark=o] table[x=power,y=rate,y expr=\thisrowno{1}*(1+\rcosrolloffa)] {results/rates_gibbs/Q8/se__senst__Gibbs__8-PAM__S=4__alpha=0.20__L=0km__mem=9__it=50__par=20__n=20000__burn=10.txt};

\addplot[ASK, line width=0.7pt,line width=0.9pt,name path global=ask8,mark=+] table[x=power,y=rate,y expr=\thisrowno{1}*(1+\rcosrolloffa)] {results/rates_gibbs/Q8/se__senst__Gibbs__8-ASK__S=4__alpha=0.20__L=0km__mem=9__it=50__par=20__n=20000__burn=10.txt};

\addplot[QAM, line width=0.7pt,line width=0.9pt,name path global=qam8,mark=x] table[x=power,y=rate,y expr=\thisrowno{1}*(1+\rcosrolloffa)] {results/rates_gibbs/Q8/se__paper__Gibbs__8-ring-qam__S=4__alpha=0.20__L=0km__mem=9__it=50__par=20__n=20000__burn=10__m=1-2__phase-offset=0__jlt_spacing=1.txt};

\addplot[PAM,line width=0.7pt,name path global=pam16,mark=square] table[x=power,y=rate,y expr=\thisrowno{1}*(1+\rcosrolloffa)] {results/rates_gibbs/Q16/se__senst__Gibbs__16-PAM__S=4__alpha=0.20__L=0km__mem=9__it=50__par=20__n=20000__burn=10.txt};

\addplot[ASK,line width=0.7pt, name path global=ask16,mark=asterisk] table[x=power,y=rate,y expr=\thisrowno{1}*(1+\rcosrolloffa)] {results/rates_gibbs/Q16/se__senst__Gibbs__16-ASK__S=4__alpha=0.20__L=0km__mem=9__it=50__par=20__n=20000__burn=10.txt};

\addplot[QAM,line width=0.7pt,name path global=qam16,mark=diamond] table[x=power,y=rate,y expr=\thisrowno{1}*(1+\rcosrolloffa)] {results/rates_gibbs/Q16/se__paper__Gibbs__16-ring-qam__S=4__alpha=0.20__L=0km__mem=9__it=50__par=20__n=20000__burn=10__m=2-2__phase-offset=1__jlt_spacing=1.txt};

\addplot[PAM,line width=0.7pt,name path global=pam32,mark=triangle] table[x=power,y=rate,y expr=\thisrowno{1}*(1+\rcosrolloffa)] {results/rates_gibbs/Q32/se__senst__Gibbs__32-PAM__S=4__alpha=0.20__L=0km__mem=9__it=50__par=20__n=20000__burn=10.txt};

\addplot[ASK,line width=0.7pt,name path global=ask32,mark=10-pointed star] table[x=power,y=rate,y expr=\thisrowno{1}*(1+\rcosrolloffa)] {results/rates_gibbs/Q32/se__senst__Gibbs__32-ASK__S=4__alpha=0.20__L=0km__mem=9__it=50__par=20__n=20000__burn=10.txt};

\addplot[QAM,line width=0.7pt,name path global=qam32,mark=*] table[x=power,y=rate,y expr=\thisrowno{1}*(1+\rcosrolloffa)] {results/rates_gibbs/Q32/se__paper__Gibbs__32-ring-qam__S=4__alpha=0.20__L=0km__mem=9__it=50__par=20__n=20000__burn=10__m=3-2__phase-offset=1__jlt_spacing=1.txt};

\addlegendentry{8-PAM}
\addlegendentry{8-ASK}
\addlegendentry{8-SQAM}
\addlegendentry{16-PAM}
\addlegendentry{16-ASK}
\addlegendentry{16-SQAM}
\addlegendentry{32-PAM}
\addlegendentry{32-ASK}
\addlegendentry{32-SQAM}

\path[name path global=line] (axis cs:\pgfkeysvalueof{/pgfplots/xmin},4) -- (axis cs: \pgfkeysvalueof{/pgfplots/xmax},4);
\path[name intersections={of=line and ask32, name=p1}, name intersections={of=line and pam32, name=p2}];
\draw[arr,thick] let \p1=(p1-1), \p2=(p2-1) in (p1-1) -- ([xshift=-0.05cm]p2-1) node [left,midway,opacitylabel,yshift=0.0cm,xshift=-0.7cm,yshift=0.1cm] {%
	\pgfplotsconvertunittocoordinate{x}{\x1}%
	\pgfplotscoordmath{x}{datascaletrafo inverse to fixed}{\pgfmathresult}%
	\edef\valueA{\pgfmathresult}%
	\pgfplotsconvertunittocoordinate{x}{\x2}%
	\pgfplotscoordmath{x}{datascaletrafo inverse to fixed}{\pgfmathresult}%
	\pgfmathparse{\pgfmathresult - \valueA}%
	\pgfmathprintnumber{\pgfmathresult} dB
};

\end{axis}
\end{tikzpicture}%
    \caption{SIC rates $I_\text{$q$,b-SIC}$ via Gibbs sampling  using FD-RC $(\alpha=0.2)$ pulses, $L=\SI{0}{\kilo\meter}$, $\widetilde{K}=9$, and $S=4$.}
    \label{fig:gibbs_results_higher}
\end{figure}
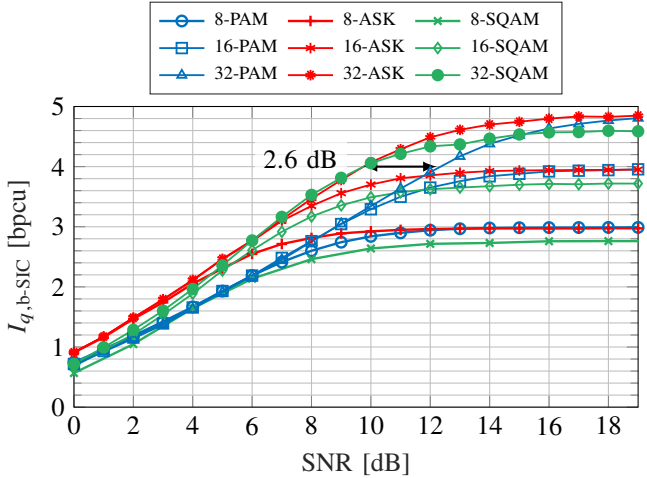

Fig.~\ref{fig:gibbs_results_higher} shows the SIC rates $I_\text{$q$,b-SIC}$ with Gibbs sampling for $\widetilde{K}=9$. Complexity prohibits using the FBA to create reference plots. ASK and PAM constellations achieve the maximum SE of $\log_2M/(1+\alpha)$ for $M=8,16$. For $M=32$, both curves saturate due to ISI; this can be resolved using $\widetilde{K}>9$. ASK gains up to $\SI{2.6}{dB}$ over PAM for $M=32$. $M$-SQAM saturates at lower rates, likely because it is difficult to distinguish the phases.
Two options to improve performance are increasing $\widetilde{K}$ and using modulations that avoid phase ambiguities. 

\begin{figure}[!t] %
    \centering
    \scalebox{0.96}{
    \begin{tikzpicture}
\renewcommand{\baselinestretch}{1}

\begin{axis}[%
MIGeneralStyle,ymax=2+0.01,
xmax=10,
xmin=-2,
ylabel={$I_{q,\text{b-SIC}}$ [bpcu]},
width=7.5cm,
ytick={0,1,2},
]

\addplot[ASK] table[x=power,y=rate] {results/rates_gibbs/Q4/senst__Gibbs__4-ASK__S=4__alpha=0.00__L=30km__mem=9__it=50__par=20__n=20000__burn=10.txt};

\addplot[ASK] table[x=power,y=rate] {results/rates_gibbs/iterations/Gibbs_iterations_it=1.txt};
\addplot[ASK] table[x=power,y=rate] {results/rates_gibbs/iterations/Gibbs_iterations_it=2.txt};
\addplot[ASK] table[x=power,y=rate] {results/rates_gibbs/iterations/Gibbs_iterations_it=5.txt};
\addplot[ASK] table[x=power,y=rate] {results/rates_gibbs/iterations/Gibbs_iterations_it=10.txt};
\addplot[ASK] table[x=power,y=rate] {results/rates_gibbs/iterations/Gibbs_iterations_it=20.txt};

\addlegendentry{4-ASK}

\draw[arrsingle,shadowed]  (axis cs:5.3,1.75) -- (axis cs:6.4,1.05) node[opacitylabel,xshift=0.1cm,below,font=black,font=\footnotesize]{$N_\mathrm{iter}=1,2,5,10,20,50$};

\end{axis}
\end{tikzpicture}%
}
    \caption{SIC rates $I_\text{$q$,b-SIC}$ via Gibbs sampling vs. number of Gibbs sampler iterations for 4-ASK, $\widetilde{K}=9$, $S=4$ and $N_\text{par} =20$ parallel samplers.}
    \label{fig:iteration_comparison}
\end{figure}
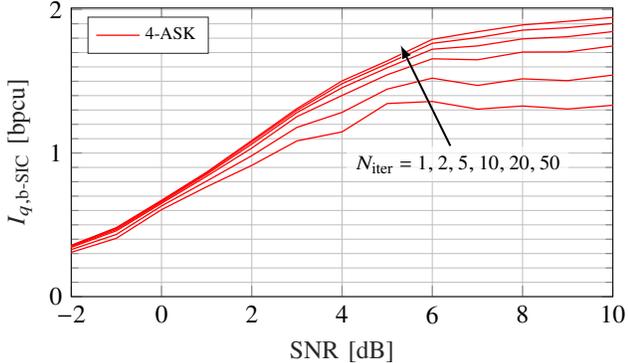

Fig.~\ref{fig:iteration_comparison} shows SIC rates for $S=4$, $\widetilde{K}=9$, and 4-ASK when using Gibbs sampling with $N_\mathrm{iter}=1,2,5,10,20,50$ iterations after the burn-in period. Observe that more iterations are needed for high SNR and that $N_\mathrm{iter}$ can be reduced at the cost of only slightly lower rates.

\subsection{Complexity and Latency}
A practical complexity comparison requires hardware considerations, which is beyond the scope of the paper. Table~\ref{tab:complexity} compares the number of multiplications required: the FBA complexity is exponential in the memory~$\widetilde{K}$, while the Gibbs sampling complexity is quadratic in~$\widetilde{K}$. Similarly, the FBA complexity is exponential in the number $m$ of bit levels since $|\mathcal{A}|= M =2^m$, while the Gibbs sampling complexity is linear in $m$. We remark that the complexity of Gibbs sampling may be reduced at intermediate SNRs; see Fig.~\ref{fig:iteration_comparison}. 

SIC increases latency by approximately a factor of $S$ as compared to SDD; see Sec.~\ref{sec:sic}. One may reduce the detection latency by precoding or choosing transmit pulses with small memory; see~\cite{tasbihi2021direct,SecondiniDirectDetectionBPAM2020}.

\section{Conclusion}\label{sec:conclusion}
Three receiver architectures were compared for DD receivers with oversampling: JDD, SDD, and SIC. SDD exhibits significant rate loss compared to JDD, especially for BM. SIC with MLC/MSD can approach JDD performance with only a few SIC stages. Simulations with polar codes showed that BM can significantly outperform intensity modulation (IM) for practical receivers.
Gibbs sampling reduces the detector complexity as compared to the FBA. Simulations at low SNR showed that a few iterations suffice to achieve rates close to the JDD rates. At high SNR, Gibbs sampling can stall.

There are many open research problems. For example, symmetric constellations exhibit phase ambiguities that motivate the use of asymmetric constellations, pulse shaping, or probabilistic constellation shaping. Similarly, one may analyze performance with noise before DD, e.g., shot noise or noise from transmitter components.

\section*{Appendix}
This appendix treats phase ambiguities for pulses symmetric about $t=0$, such as FD-RC with CD, and symbol alphabets $\mathcal{A}$ with phase symmetries, such as $M$-ASK. A similar discussion can be found in~\cite[Sec.~II]{SecondiniDirectDetectionBPAM2020}. Our main interest is high SNR, so consider the noise-free SLD output
\begin{align}
    y'(t) & = \left| \sum_\kappa X_\kappa \, g_\text{tx}(t-\kappa T_s) \right|^2
\end{align}
At $t=0$, we have
\begin{align}
   y'(0) = \left| X_0 \right|^2 \cdot |g_\text{tx}(0)|^2
   \label{eq:even-samples}
\end{align}
and similarly for all even samples; at $t=T_s/2$ we have
\begin{align}
    y'(T_s/2) & =
    \left| \sum_\kappa X_\kappa \, g_\text{tx}(T_s/2 - \kappa T_s) \right|^2 \nonumber \\
    & = \left| \sum_{\kappa\ge 1} \left( X_{\kappa} + X_{1-\kappa} \right) \, g_\text{tx}(T_s/2-\kappa T_s) \right|^2
    \label{eq:FDRC-symmetry}
\end{align}
and similarly for all odd samples. One loses the phase information in the even samples~\eqref{eq:even-samples}, and there is symmetric ISI in the odd samples. This symmetry can be problematic if $\mathcal{A}$ has a $\pm 1$ phase symmetry, for example.

To illustrate, consider BPSK with $\mathcal{A}=\{\pm 1\}$ and a mismatched receiver that accounts for only the first pulse values $g_\text{tx}(\pm T_s/2)$ in~\eqref{eq:FDRC-symmetry}.
Suppose the initial and final states are $x_0
$ and $x_{n+1}$, respectively. The resulting approximation of~\eqref{eq:FDRC-symmetry} is
\begin{align}
    y'(T_s/2) \approx
     2\left( 1 + X_0 X_1 \right) |g_\text{tx}(T_s/2)|^2
\end{align}
and we obtain the string of normalized odd samples
\begin{align}
    & \left\{\frac{y'(\kappa T_s + T_s/2)}{2 |g_\text{tx}(T_s/2)|^2} - 1 \right\}_{\kappa=0}^{n+1} %
    \approx x_0 X_1, X_1 X_2, \dots, X_n x_{n+1}.
    \label{eq:FDRC-symmetry2}
\end{align}
The expression~\eqref{eq:FDRC-symmetry2} shows that the SLD operates similarly to a differential detector on the odd samples, which explains why differential coding is useful. In fact, without differential coding, SDD is useless if $x_0=x_{n+1}=0$ because $P_{X_\kappa \lvert \mathbf{Y}}( a \lvert \mathbf{y})=1/2$ for $a=\pm 1$. Moreover, SDD is severely hampered for large $n$ if $x_0\in\{\pm 1\}$ and $x_{n+1}\in\{\pm 1\}$, since SDD must first detect $X_1$ or $X_n$, and then successively the remaining symbols.

\section*{Acknowledgment}
\noindent The authors wish to thank G.~Böcherer, T.~Rahman and  N.~Stojanovi\'{c} for helpful discussions.

\bibliographystyle{IEEEtran}
\bibliography{IEEEabrv,imdd}

\begin{thebibliography}{10}
\providecommand{\url}[1]{#1}
\csname url@samestyle\endcsname
\providecommand{\newblock}{\relax}
\providecommand{\bibinfo}[2]{#2}
\providecommand{\BIBentrySTDinterwordspacing}{\spaceskip=0pt\relax}
\providecommand{\BIBentryALTinterwordstretchfactor}{4}
\providecommand{\BIBentryALTinterwordspacing}{\spaceskip=\fontdimen2\font plus
\BIBentryALTinterwordstretchfactor\fontdimen3\font minus
  \fontdimen4\font\relax}
\providecommand{\BIBforeignlanguage}[2]{{%
\expandafter\ifx\csname l@#1\endcsname\relax
\typeout{** WARNING: IEEEtran.bst: No hyphenation pattern has been}%
\typeout{** loaded for the language `#1'. Using the pattern for}%
\typeout{** the default language instead.}%
\else
\language=\csname l@#1\endcsname
\fi
#2}}
\providecommand{\BIBdecl}{\relax}
\BIBdecl

\bibitem{chagnon_optical_comms_short_reach_2019}
M.~{Chagnon}, ``Optical communications for short reach,'' \emph{J. Lightw.
  Technol.}, vol.~37, no.~8, pp. 1779--1797, April 2019.

\bibitem{dissanayake_comparison_ofdm_imdd_2013}
S.~D. Dissanayake and J.~Armstrong, ``{C}omparison of {ACO-OFDM}, {DCO-OFDM}
  and {ADO-OFDM} in {IM/DD} systems,'' \emph{J. Lightw. Technol.}, vol.~31,
  no.~7, pp. 1063--1072, 2013.

\bibitem{chen_performance_analysis_ofdm_imdd_2012}
L.~Chen, B.~Krongold, and J.~Evans, ``Performance analysis for optical {OFDM}
  transmission in short-range {IM/DD} systems,'' \emph{J. Lightw. Technol.},
  vol.~30, no.~7, pp. 974--983, 2012.

\bibitem{mecozzi_imdd_capacity_optical_amp2001}
A.~Mecozzi and M.~Shtaif, ``On the capacity of intensity modulated systems
  using optical amplifiers,'' \emph{{IEEE} Photon. Technol. Lett.}, vol.~13,
  no.~9, pp. 1029--1031, 2001.

\bibitem{mecozzi_capacity_amdd_2018}
------, ``Information capacity of direct detection optical transmission
  systems,'' \emph{J. Lightw. Technol.}, vol.~36, no.~3, pp. 689--694, 2018.

\bibitem{SecondiniDirectDetectionBPAM2020}
M.~Secondini and E.~Forestieri, ``Direct detection of bipolar pulse amplitude
  modulation,'' \emph{J. Lightw. Technol.}, vol.~38, no.~21, pp. 5981--5990,
  2020.

\bibitem{tasbihi_capacity_waveform_channels_time-limited_2020}
A.~Tasbihi and F.~R. Kschischang, ``On the capacity of waveform channels under
  square-law detection of time-limited signals,'' \emph{{IEEE} Trans. Inf.
  Theory}, vol.~66, no.~11, pp. 6682--6687, 2020.

\bibitem{tasbihi2021direct}
------, ``Direct detection under {T}ukey signalling,'' \emph{J. Lightw.
  Technol.}, vol.~39, no.~21, pp. 6845--6857, 2021.

\bibitem{KK_receiver_mecozzi_2016}
A.~Mecozzi, C.~Antonelli, and M.~Shtaif, ``Kramers--{K}ronig coherent
  receiver,'' \emph{Optica}, vol.~3, no.~11, pp. 1220--1227, Nov 2016.

\bibitem{plabst2022achievable}
D.~Plabst, T.~Prinz, T.~Wiegart, T.~Rahman, N.~Stojanovi{\'c}, S.~Calabr{\`o},
  N.~Hanik, and G.~Kramer, ``Achievable rates for short-reach fiber-optic
  channels with direct detection,'' \emph{J. Lightw. Technol.}, vol.~40,
  no.~12, pp. 3602--3613, 2022.

\bibitem{sheik_achievable_2017}
A.~Sheikh, A.~Graell~i Amat, and G.~Liva, ``Achievable information rates for
  coded modulation with hard decision decoding for coherent fiber-optic
  systems,'' \emph{J. Lightw. Technol.}, vol.~35, no.~23, pp. 5069--5078, 2017.

\bibitem{liga_information_2017}
G.~Liga, A.~Alvarado, E.~Agrell, and P.~Bayvel, ``Information rates of
  next-generation long-haul optical fiber systems using coded modulation,''
  \emph{J. Lightw. Technol.}, vol.~35, no.~1, pp. 113--123, 2017.

\bibitem{PfisterAIRFiniteStateChan2001}
H.~Pfister, J.~Soriaga, and P.~Siegel, ``On the achievable information rates of
  finite state {ISI} channels,'' in \emph{IEEE Global Telecommun. Conf.},
  vol.~5, 2001, pp. 2992--2996 vol.5.

\bibitem{muller_capacity_separate2004}
R.~M\"{u}ller and W.~Gerstacker, ``On the capacity loss due to separation of
  detection and decoding,'' \emph{{IEEE} Trans. Inf. Theory}, vol.~50, no.~8,
  pp. 1769--1778, 2004.

\bibitem{douillard1995iterative}
C.~Douillard, M.~J{\'e}z{\'e}quel, C.~Berrou, D.~Electronique, A.~Picart,
  P.~Didier, and A.~Glavieux, ``Iterative correction of intersymbol
  interference: turbo-equalization,'' \emph{Eur. Trans. Telecommun.}, vol.~6,
  no.~5, pp. 507--511, 1995.

\bibitem{wang1999iterative}
X.~Wang and H.~V. Poor, ``Iterative (turbo) soft interference cancellation and
  decoding for coded {CDMA},'' \emph{{IEEE} Trans. Commun.}, vol.~47, no.~7,
  pp. 1046--1061, 1999.

\bibitem{ten2004design}
S.~Ten~Brink, G.~Kramer, and A.~Ashikhmin, ``Design of low-density parity-check
  codes for modulation and detection,'' \emph{{IEEE} Trans. Commun.}, vol.~52,
  no.~4, pp. 670--678, 2004.

\bibitem{wachsmann_multilevel_1999}
U.~Wachsmann, R.~F. Fischer, and J.~B. Huber, ``Multilevel codes: Theoretical
  concepts and practical design rules,'' \emph{{IEEE} Trans. Inf. Theory},
  vol.~45, no.~5, pp. 1361--1391, 1999.

\bibitem{soriaga_determining_2007}
J.~B. Soriaga, H.~D. Pfister, and P.~H. Siegel, ``Determining and approaching
  achievable rates of binary intersymbol interference channels using multistage
  decoding,'' \emph{{IEEE} Trans. Inf. Theory}, vol.~53, no.~4, pp. 1416--1429,
  2007.

\bibitem{bcjr_1974}
L.~{Bahl}, J.~{Cocke}, F.~{Jelinek}, and J.~{Raviv}, ``Optimal decoding of
  linear codes for minimizing symbol error rate,'' \emph{{IEEE} Trans. Inf.
  Theory}, vol.~20, no.~2, pp. 284--287, 1974.

\bibitem{mackay2003information}
D.~J.~C. MacKay, \emph{Information Theory, Inference and Learning
  Algorithms}.\hskip 1em plus 0.5em minus 0.4em\relax Cambridge University
  Press, 2003.

\bibitem{buchoux2000turbo}
V.~Buchoux, O.~Capp\'{e}, and {\'{E}}.~Moulines, ``Turbo multiuser detection
  for coded {DS-CDMA} systems: A {G}ibbs sampling approach,'' in \emph{Asilomar
  Conf. Signals, Sys., Computers}, vol.~2, 2000, pp. 1426--1430.

\bibitem{wang2000adaptive}
X.~Wang and R.~Chen, ``Adaptive {B}ayesian multiuser detection for synchronous
  {CDMA} with {G}aussian and impulsive noise,'' \emph{{IEEE} Trans. Signal
  Process.}, vol.~48, no.~7, pp. 2013--2028, 2000.

\bibitem{schmidl2000interference}
T.~M. Schmidl, A.~Gatherer, X.~Wang, and R.~Chen, ``Interference cancellation
  using the {G}ibbs sampler,'' in \emph{IEEE Vehic. Technol. Conf. Fall 2000},
  vol.~1, 2000, pp. 429--433.

\bibitem{shi2004markov}
Z.~Shi, H.~Zhu, and B.~Farhang-Boroujeny, ``{M}arkov chain {M}onte {C}arlo
  techniques in iterative detectors: a novel approach based on {M}onte {C}arlo
  integration,'' in \emph{IEEE Global Telecommun. Conf.}, vol.~1, 2004, pp.
  325--329.

\bibitem{yang2001turbo}
Z.~Yang and X.~Wang, ``Turbo equalization for {GMSK} signaling over multipath
  channels based on the {G}ibbs sampler,'' \emph{{IEEE} J. Sel. Areas Commun.},
  vol.~19, no.~9, pp. 1753--1763, 2001.

\bibitem{farhang2006markov}
B.~Farhang-Boroujeny, H.~Zhu, and Z.~Shi, ``{M}arkov chain {M}onte {C}arlo
  algorithms for {CDMA} and {MIMO} communication systems,'' \emph{{IEEE} Trans.
  Signal Process.}, vol.~54, no.~5, pp. 1896--1909, 2006.

\bibitem{peng2009markov}
R.-H. Peng, R.-R. Chen, and B.~Farhang-Boroujeny, ``{M}arkov chain {M}onte
  {C}arlo detectors for channels with intersymbol interference,'' \emph{{IEEE}
  Trans. Signal Process.}, vol.~58, no.~4, pp. 2206--2217, 2009.

\bibitem{peng2009low}
------, ``Low complexity {M}arkov chain {M}onte {C}arlo detector for channels
  with intersymbol interference,'' in \emph{IEEE Int. Conf. Commun.}, 2009, pp.
  1--5.

\bibitem{kashif2008monte}
F.~M. Kashif, H.~Wymeersch, and M.~Z. Win, ``{M}onte {C}arlo equalization for
  nonlinear dispersive satellite channels,'' \emph{{IEEE} J. Sel. Areas
  Commun.}, vol.~26, no.~2, pp. 245--255, 2008.

\bibitem{AgrawalFourthEdFiberOptics}
G.~Agrawal, \emph{Fiber-Optic Communication Systems}, 4th~ed.\hskip 1em plus
  0.5em minus 0.4em\relax John Wiley \& Sons, Inc., Hoboken, NJ, USA, 2010.

\bibitem{wiener_filter_plabst2020}
D.~{Plabst}, F.~J. Garc\'{i}a-{G\'{o}mez}, T.~{Wiegart}, and N.~{Hanik},
  ``Wiener filter for short-reach fiber-optic links,'' \emph{{IEEE} Commun.
  Lett.}, vol.~24, no.~11, pp. 2546--2550, 2020.

\bibitem{kavcic_binary_2003}
A.~Kav\v{c}i\'{c}, X.~Ma, and M.~Mitzenmacher, ``Binary intersymbol
  interference channels: {G}allager codes, density evolution, and code
  performance bounds,'' \emph{{IEEE} Trans. Inf. Theory}, vol.~49, no.~7, pp.
  1636--1652, 2003.

\bibitem{tal_list_2015}
I.~Tal and A.~Vardy, ``List decoding of polar codes,'' \emph{{IEEE} Trans. Inf.
  Theory}, vol.~61, no.~5, pp. 2213--2226, 2015.

\bibitem{prinz_successive_2018}
T.~Prinz and P.~Yuan, ``Successive cancellation list decoding of {BMERA} codes
  with application to higher-order modulation,'' in \emph{IEEE Int. Symp. Turbo
  Codes \& Iterative Inf. Process.}, 2018, pp. 1--5.

\bibitem{karakchieva_joint_2019}
L.~Karakchieva and P.~Trifonov, ``Joint list multistage decoding with sphere
  detection for polar coded {SCMA} systems,'' in \emph{VDE ITG Conf. Sys.,
  Commun. Coding}, 2019, pp. 1--6.

\bibitem{martinez2009bit}
A.~Martinez, A.~G. i~F\`{a}bregas, G.~Caire, and F.~M. Willems,
  ``Bit-interleaved coded modulation revisited: A mismatched decoding
  perspective,'' \emph{{IEEE} Trans. Inf. Theory}, vol.~55, no.~6, pp.
  2756--2765, 2009.

\bibitem{zhu2005performance}
H.~Zhu, B.~Farhang-Boroujeny, and R.-R. Chen, ``On performance of sphere
  decoding and {M}arkov chain {M}onte {C}arlo detection methods,'' \emph{{IEEE}
  Signal Process. Lett.}, vol.~12, no.~10, pp. 669--672, 2005.

\bibitem{mao2007markov}
X.~Mao, P.~Amini, and B.~Farhang-Boroujeny, ``{M}arkov chain {M}onte {C}arlo
  {MIMO} detection methods for high signal-to-noise ratio regimes,'' in
  \emph{IEEE Global Telecommun. Conf.}, 2007, pp. 3979--3983.

\bibitem{senst2011rao}
M.~Senst and G.~Ascheid, ``A {R}ao-{B}lackwellized {M}arkov chain {M}onte
  {C}arlo algorithm for efficient {MIMO} detection,'' in \emph{IEEE Int. Conf.
  Commun. (ICC)}, 2011, pp. 1--6.

\bibitem{hassibi2014optimized}
B.~Hassibi, M.~Hansen, A.~G. Dimakis, H.~A.~J. Alshamary, and W.~Xu,
  ``Optimized {M}arkov chain {M}onte {C}arlo for signal detection in {MIMO}
  systems: An analysis of the stationary distribution and mixing time,''
  \emph{{IEEE} Trans. Signal Process.}, vol.~62, no.~17, pp. 4436--4450, 2014.

\bibitem{hedstrom_achieving_2017}
J.~C. Hedstrom, C.~H. Yuen, R.-R. Chen, and B.~Farhang-Boroujeny, ``Achieving
  near {MAP} performance with an excited {M}arkov chain {M}onte {C}arlo {MIMO}
  detector,'' \emph{{IEEE} Trans. Wireless Commun.}, vol.~16, no.~12, pp.
  7718--7732, 2017.

\bibitem{hansen_optimal_2009}
M.~Hansen, B.~Hassibi, A.~G. Dimakis, and W.~Xu, ``Near-optimal detection in
  {MIMO} systems using {G}ibbs sampling,'' in \emph{IEEE Global Telecommun.
  Conf.}, 2009, pp. 1--6.

\bibitem{auras_vlsi_2014}
D.~Auras, U.~Deidersen, R.~Leupers, and G.~Ascheid, ``{VLSI} design of a
  parallel {MCMC}-based {MIMO} detector with multiplier-free {G}ibbs
  samplers,'' in \emph{Int. Conf. Very Large Scale Integration}, 2014, pp.
  1--6.

\bibitem{Gallager68}
R.~G. Gallager, \emph{Information Theory and Reliable Communication}.\hskip 1em
  plus 0.5em minus 0.4em\relax New York: Wiley, 1968.

\bibitem{Divsalar78}
D.~Divsalar, ``{Performance of Mismatched Receivers on Bandlimited Channels},''
  Ph.D. dissertation, Univ.\ California, Los Angeles, CA, 1978.

\bibitem{Kaplan-Shamai-A93}
G.~Kaplan and S.~Shamai~(Shitz), ``Information rates and error exponents of
  compound channels with application to antipodal signaling in a fading
  environment,'' \emph{{Archiv f\"ur Elektronik und \"Ubertragungstechnik}},
  vol.~47, no.~4, pp. 228--239, 1993.

\bibitem{Scarlett-FnT-20}
\BIBentryALTinterwordspacing
J.~Scarlett, A.~G. i~Fàbregas, A.~Somekh-Baruch, and A.~Martinez,
  ``Information-theoretic foundations of mismatched decoding,''
  \emph{Foundations and Trends® in Communications and Information Theory},
  vol.~17, no. 2–3, pp. 149--401, 2020. [Online]. Available:
  \url{http://dx.doi.org/10.1561/0100000101}
\BIBentrySTDinterwordspacing

\bibitem{Kramer-23}
G.~Kramer, ``Information rates for channels with fading, side information and
  adaptive codewords,'' \emph{Entropy}, vol.~25, no.~5, p. 728, Apr 2023.

\bibitem{arnoldsimulationmi}
D.~M. {Arnold}, H.-A. {Loeliger}, P.~O. {Vontobel}, A.~{Kav\v{c}i\'c}, and
  W.~{Zeng}, ``Simulation-based computation of information rates for channels
  with memory,'' \emph{{IEEE} Trans. Inf. Theory}, vol.~52, no.~8, pp.
  3498--3508, 2006.

\bibitem{sadeghi2009optimization}
P.~Sadeghi, P.~O. Vontobel, and R.~Shams, ``Optimization of information rate
  upper and lower bounds for channels with memory,'' \emph{{IEEE} Trans. Inf.
  Theory}, vol.~55, no.~2, pp. 663--688, 2009.

\bibitem{arikan_channel_2009}
E.~Ar{\i}kan, ``Channel polarization: A method for constructing
  capacity-achieving codes for symmetric binary-input memoryless channels,''
  \emph{{IEEE} Trans. Inf. Theory}, vol.~55, no.~7, pp. 3051--3073, 2009.

\bibitem{seidl_polar_2013}
M.~Seidl, A.~Schenk, C.~Stierstorfer, and J.~B. Huber, ``Polar-coded
  modulation,'' \emph{{IEEE} Trans. Commun.}, vol.~61, no.~10, pp. 4108--4119,
  2013.

\bibitem{bocherer2017efficient}
G.~B\"ocherer, T.~Prinz, P.~Yuan, and F.~Steiner, ``Efficient polar code
  construction for higher-order modulation,'' in \emph{IEEE Wireless Commun.
  Networking Conf. Workshops (WCNCW)}, 2017, pp. 1--6.

\bibitem{ITU-04}
``Forward error correction for high bit-rate {DWDM} submarine systems,'' Int.
  Telecommun. Union, Recommendation {ITU-T} {G.975.1}, 2004.

\bibitem{Mizuochi-09}
T.~Mizuochi, Y.~Konishi, Y.~Miyata, T.~Inoue, K.~Onohara, S.~Kametani,
  T.~Sugihara, K.~Kubo, H.~Yoshida, T.~Kobayashi, and T.~Ichikawa,
  ``Experimental demonstration of concatenated {LDPC} and {RS} codes by {FPGA}s
  emulation,'' \emph{{IEEE} Photon. Technol. Lett.}, vol.~21, no.~18, pp.
  1302--1304, 2009.

\bibitem{Magarini-10}
M.~Magarini, R.-J. Essiambre, B.~E. Basch, A.~Ashikhmin, G.~Kramer, and A.~J.
  de~Lind~van Wijngaarden, ``Concatenated coded modulation for optical
  communications systems,'' \emph{{IEEE} Photon. Technol. Lett.}, vol.~22,
  no.~16, pp. 1244--1246, 2010.

\bibitem{Smith-12}
B.~P. Smith, A.~Farhood, A.~Hunt, F.~R. Kschischang, and J.~Lodge, ``Staircase
  codes: {FEC} for 100 {G}b/s {OTN},'' \emph{J. Lightw. Technol.}, vol.~30,
  no.~1, pp. 110--117, 2012.

\bibitem{Kumar-23}
N.~Kumar, D.~Kedia, and P.~Gaurav, ``A review of channel coding schemes in the
  {5G} standard,'' \emph{Telecommun. Sys.}, vol.~83, no.~4, pp. 423--448, 2023.

\end{thebibliography}

\end{document}